\documentclass{aa}  
\usepackage{graphicx}
\usepackage{amsmath}
\usepackage{xspace}
\usepackage{booktabs}
\usepackage{lscape}
\usepackage{natbib}
\usepackage{url}
\usepackage{threeparttable}
\bibpunct{(}{)}{;}{a}{}{,}
\usepackage{txfonts}
\newcounter{Rco}

\newcommand{\Msun}{\ensuremath{{\text{M}}_\odot}\xspace}
 
\newcommand{\Teff}{\ensuremath{{T_{\text{eff}}}}\xspace}
\newcommand{\logg}{\ensuremath{{\log g}}\xspace}
\begin{document}

\title{Chandra grating spectroscopy of three hot white dwarfs}

\author{J\@. Adamczak\inst{1,2} 
        \and 
        K\@. Werner\inst{1} 
        \and 
        T\@. Rauch\inst{1} 
        \and
        S\@. Schuh\inst{3}
        \and
        J\@. J\@. Drake \inst{4}
        \and
        J\@. W\@. Kruk \inst{5}}

\institute{Institute for Astronomy and Astrophysics,
           Kepler Center for Astro and Particle Physics,
           Eberhard Karls University, 
           Sand 1,
           72076 T\"ubingen, 
           Germany           
           \and
 McDonald Observatory, The University of Texas, Austin, Texas, USA 78712,          
           \email{adamczak@astro.as.utexas.edu}
           \and 
           Institute for Astrophysics, Georg August University,
           Friedrich-Hund-Platz 1, 37077 G\"ottingen, Germany 
           \and
           Harvard-Smithsonian Center for Astrophysics, 60 Garden Street,
           Cambridge, MA 02138, USA
           \and
           NASA Goddard Space Flight Center, Greenbelt, MD 20771, USA}

\date{Received --- --- 2012 / Accepted --- --- 2012}

\abstract{
High-resolution soft X-ray spectroscopic observations of single hot white dwarfs are scarce. With
the \textsl{Chandra} Low-Energy Transmission Grating, we have observed two white
dwarfs, one is of spectral type DA (\object{LB\,1919}) and the other is a non-DA
of spectral type PG\,1159 (\object{PG\,1520+525}). The spectra of both stars are
analyzed, together with an archival \textsl{Chandra} spectrum of another DA
white dwarf (\object{GD\,246}).
}{
The soft X-ray spectra of the two DA white dwarfs are investigated in order to
study the effect of gravitational settling and radiative levitation of metals in
their photospheres. \object{LB\,1919} is of interest because it has a
significantly lower metallicity than DAs with otherwise similar atmospheric parameters. \object{GD\,246} is the only
white dwarf known that shows identifiable individual iron lines in the soft X-ray range. For
the PG\,1159 star, a precise effective temperature determination is performed in
order to confine the position of the blue edge of the GW~Vir instability region
in the HRD.
}{
The \textsl{Chandra} spectra are analyzed with chemically homogeneous
as well as stratified NLTE model atmospheres that assume equilibrium between
gravitational settling and radiative acceleration of chemical elements. Archival EUV and UV spectra obtained with
\textsl{EUVE}, \textsl{FUSE}, and \textsl{HST} are utilized to support the
analysis.
}{
No metals could be identified in  LB\,1919. All observations are compatible with
a pure hydrogen atmosphere. This is in stark contrast to the vast majority of
hot DA white dwarfs that exhibit light and heavy metals and to the stratified
models that predict significant metal abundances in the atmosphere. 
For GD\,246 we find that neither stratified nor homogeneous models can fit the \textsl{Chandra} spectrum.
The \textsl{Chandra} spectrum of PG\,1520+525 constrains the effective
temperature to 
\Teff = $150\,000 \pm 10\,000$\,K. Therefore, this nonpulsating star
together with the pulsating prototype of the GW\,Vir class (PG\,1159$-$035) defines the
location of the blue edge of the GW\,Vir
instability region. The result is in accordance with predictions from
nonadiabatic stellar pulsation models. Such models are therefore reliable tools to
investigate the interior structure of GW~Vir variables.
}{
Our soft X-ray study reveals that the understanding of
metal abundances in hot DA white dwarf atmospheres is still incomplete.  
On the other hand, model atmospheres of hydrogen-deficient PG\,1159-type
stars are reliable and reproduce well the observed spectra from soft X-ray
to optical wavelengths. 
}


\maketitle

\section{Introduction \label{sect:intro}}

An unexpectedly small number of white dwarfs (WDs) were detected in the
\textsl{ROSAT} PSPC X-ray all sky survey \citep{1996A&A...316..147F}. Most of
them (161) are of spectral type DA (pure-hydrogen optical spectra), with an
additional three DOs (helium-dominated), three DAOs (mixed H/He optical
spectra), and eight PG\,1159 stars (He--C--O dominated). It was realized that
the atmospheric opacity of radiatively levitated metals effectively blocks the
outward leakage of X-rays in the vast majority of DAs with \Teff $>$ 40\,000\,K
\citep{1997MNRAS.286...58B}. Furthermore, the interstellar medium proved
denser than expected in many lines of sight (see,
e.g., \citealt{1999A&A...352..308W}), thus affecting cooler WDs.

High-resolution soft X-ray spectroscopy of hot WDs enables the identification of
chemical species in their photospheres that cannot be detected in other
wavelength ranges. However, such observations are rather scarce and only became
feasible with the advent of the \textsl{Chandra} observatory with its Low-Energy
Transmission Grating (LETG).

Until recently, only five WDs were observed with \textsl{Chandra}
LETG. \object{H\,1504+65} is an extremely hot (\Teff= 200\,000\,K), peculiar
PG\,1159 star, obviously a naked C--O or O--Ne WD, whose soft X-ray spectrum is
characterized by highly ionized O, Ne, and Mg lines
\citep{2004A&A...421.1169W}. On the other hand, \object{GD\,246} is a DA, and
its \textsl{Chandra} spectrum allowed the first unambiguous identification of
lines from highly ionized iron \citep{2002ASPC..262...57V}. The \textsl{Chandra}
spectra of the two DAs \object{HZ\,43} and \object{Sirius~B} are perfectly
matched by pure-hydrogen atmospheres and are used as soft-X-ray calibration
targets \citep{2000SPIE.4012..700P,2002ASPC..262...57V,2006A&A...458..541B}. A
\textsl{Chandra} observation of the hot DO \object{KPD\,0005+5106} (\Teff=
200\,000\,K) was used to prove that the soft X-ray emission is of photospheric
and not coronal origin; however, the S/N ratio is insufficient to identify
individual spectral lines \citep{2005ApJ...625..973D}.

Few other WDs are bright enough in soft X-rays to obtain useful
\textsl{Chandra} spectra. We have observed two of them (\object{LB\,1919} and
\object{PG\,1520+525}) and including archival data of
\object{GD\,246} present a spectral analysis in this paper. The DA
\object{LB\,1919} is of interest because previous \textsl{Extreme Ultraviolet
Explorer (EUVE)} spectroscopy showed an unexpectedly high soft X-ray flux that
reveals an unexplained low metal abundance \citep{1998AuAfortable...329.1045W},
in contrast to expectations from radiative-levitation theory. We performed X-ray
spectroscopy in order to identify individual elements that could hint at the
origin of the metal deficiency.

We observed \object{PG\,1520+525} in order to constrain its effective
temperature. The motivation lies in the fact that this nonpulsating PG\,1159
star, together with the pulsating prototype \object{PG\,1159--035}, defines the
blue edge of the GW\,Vir instability region in the Hertzsprung-Russell diagram
(HRD), see e.g. \cite{2007A&A...462..281J}. The homogeneous atmosphere of
\object{PG\,1520+525} provides a convenient comparison for the investigated DA
white dwarfs because it is not subject to the uncertainties in the treatment of
radiative levitation and stratification of atmospheric composition. Thus,
\object{PG\,1520+525} provides a test case for the model atmospheres employed in this
study.  A failure to achieve a consistent fit to the UV/optical and X-ray
spectra of the homogeneous PG1159 photosphere would indicate serious model
deficiencies in addition to problems that would be encountered with radiative
levitation physics in the case of hot DA WDs.

This paper is organized as follows. We first describe in more detail the
motivation of our soft X-ray analyses (Sect.\,\ref{sect:softx}). Then we specify
our model atmosphere calculations and the atomic data used
(Sect.\,\ref{sect:models}). In Sect.\,\ref{sect:obs}, we report on the
observations utilized in our analysis. In Sections \ref{sect:lb} to
\ref{sect:pg} we delineate the analysis procedure of our three program stars one
at a time, and in Sect.\,\ref{sect:sum} we conclude.

\section{Soft X-ray emission from white dwarfs}\label{sect:softx}

A concise presentation of past soft X-ray observations of WDs and their
interpretation can be found, e.g., in \citet{we:08}.

\subsection{DA white dwarfs}\label{sect:das}

DA WDs possess almost pure hydrogen atmospheres. The high gravity in these
objects results in a chemical stratification with the lightest element,
hydrogen, floating atop. In hot DAs (\Teff $>$ 20\,000\,K), hydrogen is almost
completely ionized and the opacity in the atmosphere is strongly reduced. Soft
X-ray radiation can emerge from deep, hot, photospheric layers.  However, this
radiation may be blocked by the opacity of heavy elements that can be kept in
the atmospheres by radiative levitation \citep{1995ApJS...99..189C,
1995ApJ...454..429C}. The increasing efficiency of this mechanism with \Teff
results in only a few X-ray detected DAs with \Teff$>60\,000\,\text{K}$.

Since the spectral resolution of \textsl{EUVE} is too low to identify lines of
individual species, the metal abundances of the hitherto investigated DAs were
determined relative to the well-studied \object{G\,191--B2B}
\citep{1998AuAfortable...329.1045W} using chemically homogeneous model
atmospheres. The scaling of the metallicity in these models relative to that of
\object{G\,191--B2B} (the so-called metallicity index) results in satisfying
spectral energy distribution (SED) fits for many, but not all, DAs.

To make further progress,  non-LTE model atmospheres were developed that
calculate the abundances of the elements at each atmospheric depth in a
self-consistent way, assuming equilibrium between gravitational downward pull
and radiative upward acceleration \citep{1999A&A...352..632D}. The \textsl{EUVE}
SEDs of most DAs can be reproduced well by these models
\citep{2002AuAfortable...382..164S}. Some stars show a higher metallicity than
predicted. This can be explained, for instance, by ongoing accretion of
circumstellar or interstellar matter. Other stars have a lower metallicity than
predicted by these models, and the reason for this is not known. A drastic
example is \object{HZ\,43A} ($\Teff = 51\,000$\,K,
\citealt{2006A&A...458..541B}) whose SED is perfectly matched by a pure H model
atmosphere. Another low-metallicity DA is the even hotter \object{LB\,1919}
($\Teff = 56\,000$\,K, this paper), which is subjected to analysis here.

The exact knowledge of element abundances in the atmospheres of the metal-poor
DAs might give hints as to an explanation for their metal deficiency. Many
spectral lines of the high ionization stages of heavy elements like iron and
nickel lie in the soft X-ray wavelength range. Spectroscopy with
\textsl{Chandra} enables us to investigate these objects more closely. Our work
represents the first detailed analysis of high-resolution soft X-ray spectra of
DAs. Besides \object{LB\,1919}, we study the more metal-rich DA \object{GD\,246}
as a reference object.

\begin{figure}
\centering \rotatebox{-90}{\includegraphics[scale=0.8]{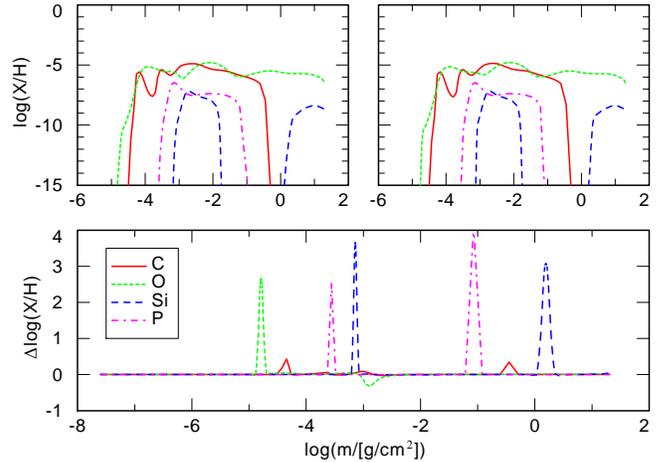}}
  \caption{Top panels: Element abundances over depth following from two different
    treatments of radiative acceleration by photoionization
    ($\Teff=54\,000$\,K, $\logg=8.2$). Left: the whole photon momentum is
    transferred to the ion.  Right: only a fraction is
    transferred, according to Eq.\,\ref{fionSommerfeld}. Bottom
    panel: difference. The peaks correspond to a shift of the steep regions of
    the abundance pattern by one depth point in the model atmosphere.}
\label{Bound-Free_alt_neu_HCOSiP}
\end{figure}

\subsection{PG\,1159 stars}\label{subsect:pgstars}

As mentioned in the introduction, only very few non-DAs (DO and PG\,1159 stars)
were detected in the soft X-ray band because of the additional opacity of
helium and enriched metals. PG\,1159 stars are hot, H-deficient (pre-) WDs with
atmospheres mainly composed of He, C, and O. It is thought that their H-envelope
was consumed during a late He-shell flash (\citealt{2006PASP..118..183W}, and
references therein). Among the PG\,1159 stars, \object{PG\,1520+525} (\Teff =
150\,000\,K) is one of the brightest soft X-ray sources and hence an
interesting target to study the spectral characteristics of this class in that
wavelength region. The surface abundances of these hot stars are not
affected by gravitational and radiative acceleration  because of a weak
radiation-driven wind \citep{1997A&A...321..485U}. Consequently, their
atmospheres and envelopes can be assumed to be chemically homogeneous.

\begin{figure*}
\centering
\rotatebox{90}{\includegraphics[width=0.40\linewidth]{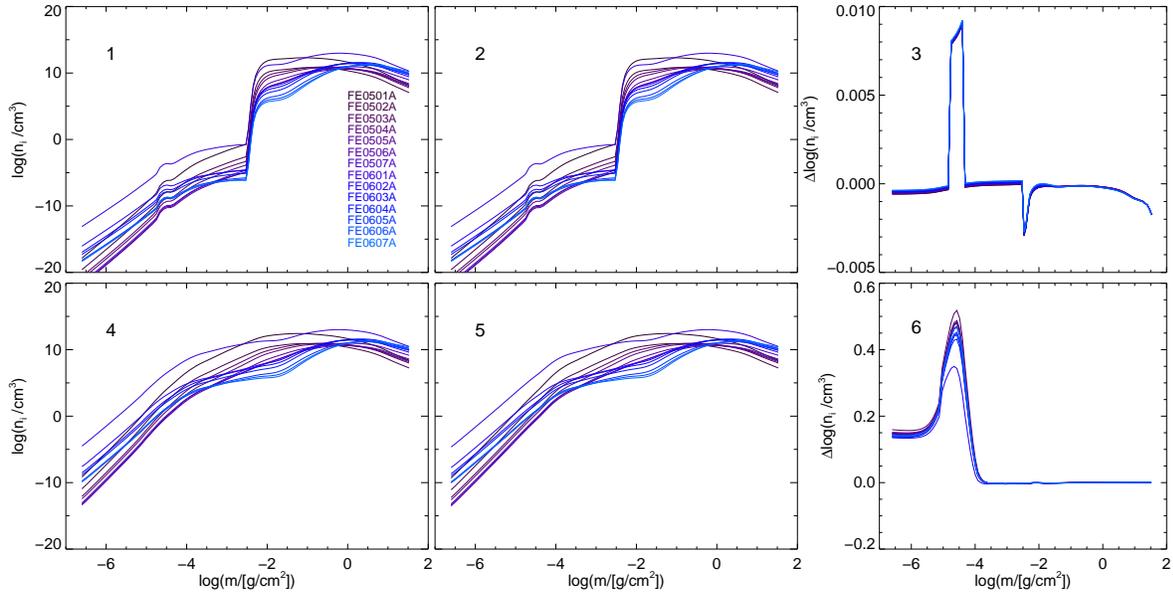}}
\vspace{5mm}
\caption{Effect of modified treatments of acceleration by line and
  photoionization transitions on occupation numbers of Fe\,{\sc v} and Fe\,{\sc
  vi} superlevels  ($\Teff=56\,000$\,K, $\logg=7.9$). The six panels show:
  1. Original treatment of bb and bf
  transitions. 2. New treatment of bf transitions. 3. Difference between the
  first two. 4. New procedure for bb transitions. 5. bf and bb refinements are
  implemented. 6. Difference between the last
  two. The legend in panel~1 refers to the superlevel designations.\label{Bound-Free_Bound-Bound_change_iron}}
\end{figure*}

Our primary aim for the \textsl{Chandra} observation of the nonvariable
\object{PG\,1520+525} is to constrain its effective temperature. Then we can
compare its position in the log\,\Teff--\logg\ diagram with that of the
prototype of the GW~Vir pulsators, PG\,1159--035. The blue edge of the
instability region is confined by these stars \citep{1996aeu..conf..229W}. Based
on nonadiabatic computations, the most advanced pulsation models for GW~Vir
stars, are those presented by \citet{0067-0049-171-1-219}. Accordingly, the
exact location of the blue edge depends on the envelope composition, primarily
on the C and O abundance and, to a smaller extent, on the
metallicity. Therefore, the concept of a blue edge is necessarily ``fuzzy''. It
was shown by spectroscopic analyses that both stars have (within error limits)
the same abundances of C, O, and Fe (see below) and thus define the blue edge
for that particular composition. We can therefore prove or disprove the
prediction of corresponding pulsation models as to the location of the
edge. This represents a strong test for these models and their potential to
derive the interior structure of PG\,1159 stars by asteroseismology methods.

\section{Model atmospheres and atomic data}\label{sect:models}

For the spectral analysis, chemically homogeneous NLTE model atmospheres were
computed with the T\"ubingen Model Atmosphere Package
(\emph{TMAP}\footnote{\url{http://astro.uni-tuebingen.de/~TMAP}}, 
\citealt{2003ASPC..288...31W}), and chemically stratified models with a special
variant, the diffusion/levitation NGRT code \citep{1999A&A...352..632D}. In
comparison to earlier work with NGRT, refinements to the code were applied for
more realistic physics. Moreover, new model atoms for Fe, Ni, and Ge were
constructed. These improvements are described in the following.

\subsection{Improvements to the diffusion/levitation code NGRT}

In the NGRT code, the element abundance pattern at any depth point in the
atmosphere is calculated self-consistently by assuming equilibrium between
gravitational and radiative forces. As a result, the atmosphere is no longer
chemically homogeneous but vertically stratified.

\subsubsection{Bound-free transitions}

The primary assumption for the calculation of the radiative force of a
bound-free (bf) transition in NGRT is that the photon momentum is transferred
completely to the ion, keeping the electron out of consideration. As a
consequence, the calculated radiative acceleration of the remaining ion might be
overestimated. To check this assumption quantitatively, a correction according
to \citet{1995A&A...297..223G} was implemented and tested.

The radiative force on a particle of element $A$, with ionic charge i for a
bound-bound (bb) or bf transition j, caused by photons in the frequency range
($\nu, \nu+d\nu$) is
\begin{align}
F_{\nu}^{\text{ij}}d\nu=\frac{dp_{\nu}}{dt}d\nu=\sigma_{\text{ij}}(\nu)\frac{\mathcal{F_{\nu}}}{c}d\nu,
\end{align}
with the net momentum $dp_{\nu}$, transported by the radiation flux
$\mathcal{F_{\nu}}$.  As mentioned above, in the case of a bf
transition, the momentum of the photon causing an ionization of ion
$A^{\text{i}}$ is not completely transferred to ion $A^{\text{i+1}}$. A part of
it is taken away by the ejected electron. Thus, a correction factor
$f_\text{ion}$ has to be introduced, describing the remaining fraction of the momentum
transferred to the ion $A^{\text{i+1}}$.

The correction factor can be expressed in terms of the frequency $\nu$ of the
photon, the threshold frequency $\nu_0$ necessary to eject the electron and a
factor $a_1$.
\begin{align}
f_\text{ion}=1-\frac{4}{3}\frac{\left(\nu-\nu_0\right)}{\nu}a_1.
\end{align}
For testing purposes, the quantum calculation of \citet{1995Sommerfeld} was taken
and applied to the code. With his value of $a_1=6/5$, $f_\text{ion}$ becomes
\begin{align}\label{fionSommerfeld}
f_\text{ion}=1-\frac{8}{5}\frac{h\nu-\chi}{h\nu},
\end{align}
with the ionization energy
threshold $\chi$.  An interesting result of Eq. \ref{fionSommerfeld} was already
stated by \citet{1970ApJ...160..641M}. For certain frequencies the electron can
be ejected with more momentum than brought in by the photon
($f_{\text{elec}}=1-f_\text{ion}>1$). As a consequence, the ion will be pushed
back in the atmosphere by the photoionization. Thus, including a detailed
bf absorption calculation can in principle lead to either higher or lower
outward directed radiative forces on an ion.

The comparison between abundances at each depth of the atmospheres calculated
with old and new treatments of bf transitions reveals, however, only minor and practically
unimportant deviations (Fig.\,\ref{Bound-Free_alt_neu_HCOSiP}); there is no
effect on the SED, in particular on the spectral lines, of the model.

\subsubsection{Bound-bound transitions}

In the original NGRT code as used by \cite{2002AuAfortable...382..164S}, the
redistribution of transferred photon momentum over the ionization stages
$A^{\text{i}}$ and $A^{\text{i+1}}$ by a bb transition in
$A^{\text{i}}$ was treated such that the momentum was completely transferred to
the next higher ion $A^{\text{i+1}}$. This treatment is based on the assumption
that an ionization following the radiative excitation of the ion takes place
before the ion is deexcited by an inelastic electron collision. Especially in
very hot atmospheres, this need not be true. For a realistic treatment, the
probabilities for ionization and collisional deexcitation would have to be
evaluated. In a statistical sense, fractions of the momentum are transferred to
the next higher ion $A^{\text{i+1}}$ and to the originally excited ion
$A^{\text{i}}$.
 
As a rough approach to test the maximum possible effect, the calculation of the
bb transitions was modified such that the entire photon momentum is transferred
to the \textsl{lower} ion $A^{\text{i}}$. For example, we show the effect of the
modified bb (and bf) procedures on the population numbers of iron superlevels
(introduced below) in Fig.\,\ref{Bound-Free_Bound-Bound_change_iron}. While the
modified bf transitions leave the occupation numbers essentially unaltered, the
modification of the bb transitions results in an obvious change, although only
in the outer photospheric layers. The effect on the emergent flux is below 10\%
and thus considered insignificant, even under the extreme assumption of the
fully inverted momentum distribution. Therefore, a more complex treatment is
unnecessary for our purposes.

\begin{figure}
\centering
\includegraphics[width=0.7 \linewidth]{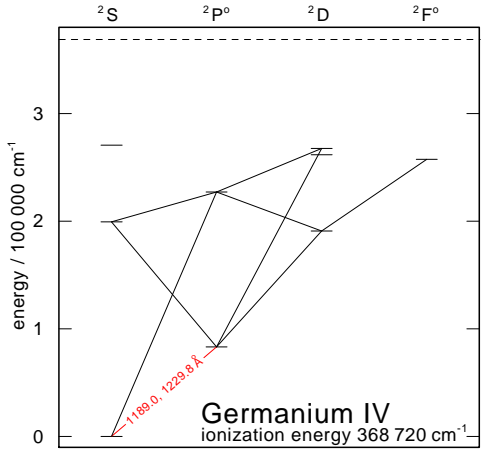}
\caption{Grotrian diagram of Ge\,{\sc iv}. The resonance doublet is
  labeled. \label{grotrian_GE_IV}}    
\end{figure} 

\begin{figure}
\centering \rotatebox{-90}{\includegraphics[width=0.40\linewidth]{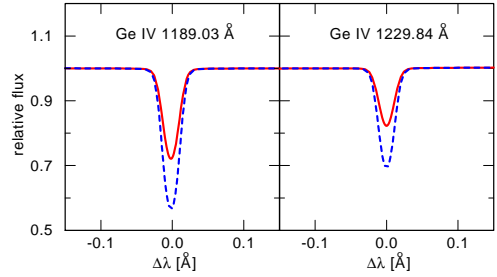}}
\caption{Ge\,IV line profiles from LTE (dashed) and NLTE (solid) H+Ge model
  atmospheres ($\Teff=57\,000$\,K,
  $\logg=7.6$, Ge/H $=2.5 \cdot 10^{-9}$). \label{GD246_Ge_lte_nlte}}
\end{figure} 

\subsection{Model atoms}\label{sect:adata}  

For our model atmosphere calculations, we used two different sets of model atoms
for the analysis of the PG\,1159 star on the one hand and for the two DAs on the other
hand. The first set is identical to that described in detail by
\citet{2004A&A...421.1169W} in their \textsl{Chandra} analysis of H\,1504+65. It
comprises He, C, O, Ne, and Mg.  The second set incorporates H, He, C, O, Si, P,
S, and Ge, being represented by detailed model atoms taken from the
T\"ubingen Model Atom
Database\footnote{\url{http://astro.uni-tuebingen.de/~TMAD}} \emph{TMAD}.
It is augmented by Fe and Ni which are treated in a
statistical way using superlevels and superlines
\citep{2003ASPC..288..103R.neu}. Table~\ref{WDmodelatoms} summarizes the set of
data used for the DA analyses.

The Ge model atom is new and presented in more detail below. We also
use new Fe and Ni model atoms that comprise vastly extended line lists \citep{Kuruczonline}
compared to all previous analyses. Their characteristics are also described in
detail below.

\begin{table}
\begin{center}
\begin{threeparttable}
\small
\caption{Statistics of model atoms used in the calculations for \object{LB\,1919} and
  \object{GD\,246}. Numbers in brackets denote levels and lines combined into superlevels
  and superlines. 
\label{WDmodelatoms}}
\begin{tabular}{llrrrr}
\toprule
Elem.         & Ion  & NLTE levels & & lines &  \\
\midrule
H             & I    & 10          &                   & 45    &           \\
              & II   & 1           &                   & $-$   &           \\
C             & II   & 16          &                   & 37    &           \\
              & III  & 58          &                   & 329   &           \\
              & IV   & 54          &                   & 295   &           \\
              & V    & 1           &                   & $-$   &           \\
O             & II   & 16          &                   & 26    &           \\
              & III  & 21          &                   & 38    &           \\
              & IV   & 18          &                   & 39    &           \\
              & V    & 30          &                   & 95    &           \\
              & VI   & 54          &                   & 291   &           \\
              & VII  & 1           &                   & $-$   &           \\
Si   & II \tnote{a}  & 20          &                   & 48    &           \\
              & III  & 17          &                   & 28    &           \\
              & IV   & 16          &                   & 44    &           \\
              & V    & 15          &                   & 20    &           \\
              & VI   & 1           &                   & $-$   &           \\
P             & III  & 3           &                   & 0     &           \\
              & IV   & 15          &                   & 9     &           \\
              & V    & 18          &                   & 12    &           \\
              & VI   & 1           &                   & $-$   &           \\
S   & III \tnote{a}  & 3           &                   & 0     &           \\
              & IV   & 17          &                   & 32    &           \\
              & V    & 19          &                   & 32    &           \\
              & VI   & 18          &                   & 48    &           \\
              & VII  & 1           &                   & $-$   &           \\
Ge \tnote{b}  & III  & 14          &                   & 0     &           \\
              & IV   & 8           &                   & 8     &           \\
              & V    & 9           &                   & 0     &           \\
              & VI   & 1           &                   & $-$   &           \\ 
Fe            & III  & 7           &  (2467)           & 27    & (537689)  \\
              & IV   & 7           &  (6389)           & 27    & (3102371) \\
              & V    & 7           &  (6728)           & 25    & (3266247) \\
              & VI   & 8           &  (5464)           & 33    & (991935)  \\
              & VII  & 9           &  (2690)           & 41    & (200455)  \\
              & VIII & 8           &  (694)            & 33    & (19587)   \\
              & IX   & 1           &                   & $-$   &           \\
Ni            & III  & 5           &  (3739)           & 14    & (1033920) \\
              & IV   & 7           &  (5939)           & 27    & (2512561) \\
              & V    & 7           &  (6165)           & 25    & (2766664) \\
              & VI   & 7           &  (10985)          & 27    & (7408657) \\
              & VII  & 8           &  (11866)          & 33    & (4195381) \\
              & VIII & 7           &  (6469)           & 27    & (1473122) \\
              & IX   & 1           &                   & $-$   &           \\
\bottomrule
\end{tabular}
\begin{tablenotes}[para]\footnotesize
\item[a] LB\,1919 models only
\item[b] GD\,246 models only
\end{tablenotes}
\normalsize
\end{threeparttable}
\end{center}
\end{table}

\begin{figure}
\centering \rotatebox{90}{\includegraphics[width=0.7\linewidth]{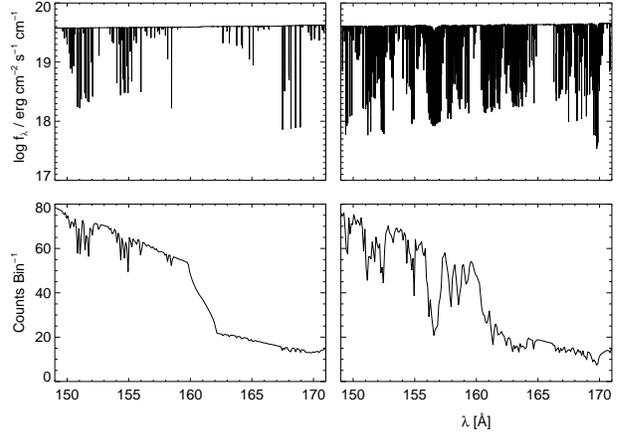}}
\caption{Comparison of H+Fe model spectra calculated with old and new Kurucz line
  lists (top left and right panels, respectively; $\Teff=57\,000$\,K,
  $\logg=7.90$, Fe/H = $1.25\cdot 10^{-6}$). Bottom panels: same models after
  convolution with the \textsl{Chandra} LETG instrument response (bin size
  0.1\,\AA). \label{Kurucz_old_new_vgl_spek}}
\end{figure}

\subsubsection{Germanium\label{subsectionGemodelatom}}

Germanium (Ge\,{\sc iv}) was first identified in three hot DAs, among them
GD\,246, by \citet{2005ApJ...622L.121V}.  For GD\,246, we perform the first NLTE
analysis of Ge in a stellar atmosphere and investigate its radiative
levitation properties. The model atom was constructed using level energies from
the NIST\footnote{\url{http://physics.nist.gov/}} database. Due to the lack of
data, all oscillator strengths $f_{\rm ij}$ are approximated by adopting values
from the isoelectronic C\,{\sc iv}, with the exception of the two observed
Ge\,{\sc iv} $\lambda\lambda\,1189.07, 1229.84$\,\AA\ lines, for which we use
the $f_{\rm ij}$ from \citet{2005ApJ...630L.169C}. Very recently, new $f_{\rm ij}$
for Ge\,{\sc iv} lines were published \citep{2011ApJ...737...25N}. A single test
calculation was performed in which we use the new data for all line transitions
in the model atom. The effect on the observed two lines is insignificant. 
Photoionization rates were computed with hydrogen-like cross sections. Electron
collisional excitation and ionization rates were evaluated with usual
approximation formulae following \citet{1962ApJ...136..906V} and
\citet{1962amp..conf..375S}, respectively. 

In
order to check for the importance of NLTE effects, we performed another test
calculation in which we enforced LTE populations for Ge levels by artificially
increasing electron collisional rates. The NLTE line profiles are significantly
shallower than the LTE profiles (Fig.\,\ref{GD246_Ge_lte_nlte}). As a
consequence, the derived NLTE abundances will be roughly a factor of two higher
compared to LTE analyses.

\subsubsection{Iron and nickel}

\begin{figure}
\centering \rotatebox{90}{\includegraphics[width=0.7\linewidth]{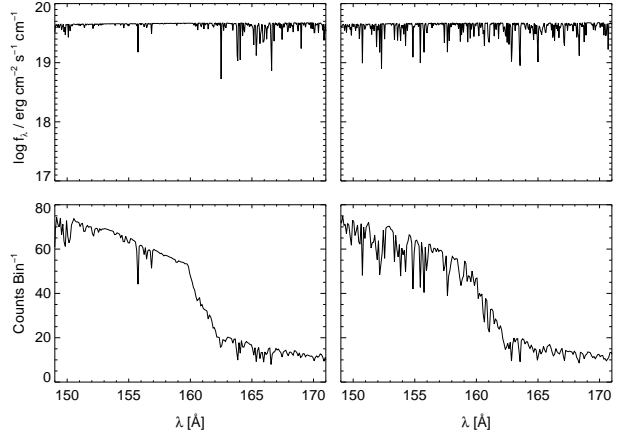}}
\caption{Like Fig.\,\ref{Kurucz_old_new_vgl_spek}, but here for nickel 
(Ni/H = $7.9\cdot 10^{-8}$). \label{Paper_newNi_lin_neu}}
\end{figure}

The \textsl{Iron Opacity Interface} (I{\sc r}O{\sc n}I{\sc c}) constructs model
atoms for iron-group elements as input for the stellar-atmosphere code
\citep{2003ASPC..288..103R.neu}. During the course of this work, new Fe and
Ni line data were provided by
\citet{Kuruczonline}\footnote{\url{http://kurucz.harvard.edu/atoms.html}}. They
include a much larger number of lines (about 20 times more), especially in the X-ray
range. Figures~\ref{Kurucz_old_new_vgl_spek} and \ref{Paper_newNi_lin_neu}
demonstrate the effect of these additional lines used in the atmosphere
calculations. The new line lists were not incorporated in the bulk of the model
calculations, however, the best-fitting models presented here were recalculated
with the newest Kurucz atomic data sets available.

\section{Observations, reddening, interstellar H and He column densities}\label{sect:obs} 

We performed \textsl{Chandra} spectroscopy of \object{LB\,1919} and
\object{PG\,1520+525} with the LETG/HRC-S. For
\object{GD\,246}, we used the \textsl{Chandra} observation investigated by
\citet{2002ASPC..262...57V}. Other observations used in our work were retrieved
from the MAST\footnote{\url{http://archive.stsci.edu/}} archive. All
observations are listed in Table~\,\ref{Observationtable}.

\begin{table}
\begin{center}
\small
\caption{Observation log of spectra analyzed in our study. \label{Observationtable}}    
\begin{tabular}{lllr}
\toprule
Star                  & Instrument (Id.)& Observation & $t_{\text{exp}}$ \\
                      &           & start date (UT)               & (ks)\\
                      &           & mm-dd-yyyy& \\
\midrule
\object{LB\,1919}     & \textsl{EUVE}         & 04-16-1994 & 166 \\
                      & \textsl{FUSE} LWRS    & 05-03-2001 &  33 \\
                      & \textsl{IUE} SWP 52808     & 11-15-1994 & 4.5 \\
                      & \textsl{Chandra} LETG/HRC-S & 02-02-2006 & 111 \\
\noalign{\smallskip}
\object{GD\,246}      & \textsl{EUVE}         & 08-08-1994 &  14 \\
                      & \textsl{FUSE} LWRS    & 11-12-2000 & 1.5 \\
                      & \textsl{FUSE} MDRS    & 07-14-2001 &  24 \\
                      & \textsl{HST} STIS/E140H     & 11-20-1998 & 2.4 \\
                      & \textsl{IUE} SWP 40467     & 12-27-1990 & 0.33 \\
                      & \textsl{IUE} LWP 19485      & 12-27-1990 & 0.84 \\
                      & \textsl{Chandra} LETG/HRC-S & 01-14-2000 &  40 \\
\noalign{\smallskip}
\object{PG\,1520+525} & \textsl{Chandra} LETG/HRC-S & 04-04-2006 & 142 \\
\bottomrule
\end{tabular}
\normalsize
\end{center}
\end{table}
 
\begin{figure}
\centering \rotatebox{-90}{\includegraphics[width=0.7\columnwidth]{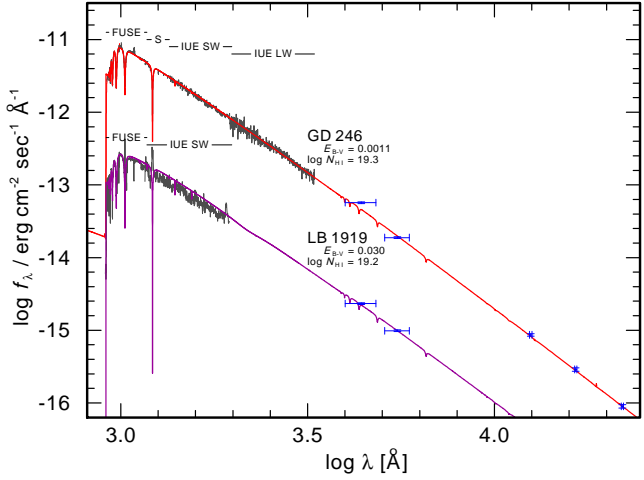}}
\caption{Determination of interstellar reddening of LB\,1919 and
GD\,246 from the observed flux distribution compared to models (thick red lines). For clarity, all
spectra are convolved with 2\,\AA\ (FWHM) Gaussians.\label{EBV_FUSE_IUE}}
\end{figure}

In order to determine interstellar reddening, \textsl{FUSE}, \textsl{IUE}, and
\textsl{HST}-STIS spectra (the latter marked with ``S'' in
Fig.\,\ref{EBV_FUSE_IUE}) as well as optical and IR magnitudes from the SIMBAD and
NOMAD databases were investigated. Fluxes of our final models were scaled to the
reddest available photometric magnitude. The reddening law of
\citet{1999PASP..111...63F} was used with $R_v=3.1$.  We determined
$E_{\text{B}-\text{V}}=0.030\pm 0.005$ and $0.0011\pm 0.0001$ for LB\,1919 and
GD\,246, respectively.  These quantities will be used below for distance
determination.  Accounting for absolute-flux uncertainties in the UV, more
realistic errors are of the order $\pm 0.05$. However, the error in the distance
determination is dominated by the uncertainties in \Teff and \logg, and we
neglect the errors for reddening. For PG\,1520+525,
\citet{1998A&A...334..618D} derived $E_{\text{B}-\text{V}}=0.0$.

Column densities for interstellar neutral hydrogen as well as neutral and
ionized helium affect the analysis of \textsl{EUVE} and \textsl{Chandra}
spectra.

LB\,1919: Based on the values found in the \textsl{EUVE} analysis by
\citet{2005PhDT.........1L} ($\text{N(H\,{\sc i})}=1.6 \cdot
10^{19}\text{cm}^{-2}$,  $\text{N(He\,{\sc i})}=0.04\,\text{N(H\,{\sc i})}$,
$\text{N(He\,{\sc ii})}=0.04\,\text{N(H\,{\sc i})}$), the interstellar H and He
column densities were varied in our fitting procedure. Our values for two of the
best-fitting models are depicted in Fig.\,\ref{LB1919_euve_bestfit}. For our
present fit to the \textsl{Chandra} spectrum, we used $\text{N(H\,{\sc i})}=1.6
\cdot 10^{19}\text{cm}^{-2}$,  $\text{N(He\,{\sc i})}=0.02\cdot\text{N(H\,{\sc
i})}$, $\text{N(He\,{\sc ii})}=0.04\cdot\text{N(H\,{\sc i})}$.

GD\,246: The H\,{\sc i} column density of $1.288 \cdot 10^{19} \text{cm}^{-2}$ was
taken from \cite{2003ApJ...587..235O} and kept fixed. 
Our adopted values for the column densities of He\,{\sc i} (0.12 $\text{N(H\,{\sc
i})}$) and He\,{\sc ii} (0.02 $\text{N(H\,{\sc
i})}$) are close to the ones determined by
\citet{1993ApJ...410L.119V} ($\text{N(He\,{\sc i})}=1.05-1.25 \cdot
10^{18}\text{cm}^{-2}$, $\text{N(He\,{\sc ii})}=3.4-4.0 \cdot
10^{17}\text{cm}^{-2}$).

Since these numbers depend on their assumed
photospheric model, we adjusted the values slightly to better fit our
models. For our \textsl{Chandra} analysis presented here we chose 0.05 and
0.026\,N(H\,{\sc i}), respectively, which is close to the values used by
\citet{2005PhDT.........1L}.

PG\,1520+525:  \citet{1996aeu..conf..229W} used $\text{N(H\,{\sc i})}=1.5\cdot
10^{20}\,\text{cm}^{-2}$ for their fit of a \Teff = 150\,000\,K model to the SED
observed by \textsl{EUVE}.  \citet{1998A&A...334..618D} determined
$\text{N(H\,{\sc i})}=2.5\cdot 10^{20}\,\text{cm}^{-2}$ from the  Ly$\alpha$
line profile. For the fit to the \textsl{Chandra} spectrum, the H\,{\sc i}
column density was treated as a free parameter. For our final model fits in this
paper, we found $1.0\cdot 10^{20}\,\text{cm}^{-2}$. The variation of the He
column densities had no effect on the fit quality, and were kept fixed at
$\text{N(He\,{\sc i})}=0.05\,\text{N(H\,{\sc i})}$ and $\text{N(He\,{\sc
ii})}=0.026\cdot\text{N(H\,{\sc i})}$.

\section{\object{LB\,1919}}\label{sect:lb} 

\subsection{Previous investigations \label{previnvestLB1919}}

\object{LB\,1919} (\object{WD\,1056+516}) was  analyzed by
\citet{1997ApJ...480..714V} using Balmer-line spectroscopy and pure-hydrogen
model atmospheres. They obtained $\Teff=68\,640$\,K and $\logg=8.08$ (cgs
units). \citet{1997ApJ...488..375F} excluded the star from their analysis. Their
spectrum exhibited flat-bottomed Balmer-line profiles, and they suggested orbital
or rotational velocities of $\sim 1000$\,km/s as a possible
interpretation, although they may be artifacts (Koester priv. comm.). 
More recently, \citet{2007ApJ...667.1126L} determined
$\Teff=67\,022\,\text{K}$ and $\logg=7.99$, and \citet{2011ApJ...743..138G}
found $\Teff=68\,510\,\text{K}$ and $\logg=7.94$ by pure-hydrogen model fits to
the Balmer lines.

\citet{1998AuAfortable...329.1045W} investigated a sample of 20 DAs  from the
\textsl{EUVE} archive. Introducing the above-mentioned metallicity index
(Sect.\,\ref{sect:das}), they were able to reproduce the spectra of most of
them. For \object{LB\,1919}, they determined a metallicity at least ten
times smaller compared to that of \object{G\,191--B2B}, adopting \Teff and \logg
values from \citet{1997ApJ...480..714V}.

Based on these results, \citet{2005PhDT.........1L} performed an analysis of
\textsl{EUVE} spectra of DAs using new chemically stratified NLTE model
atmospheres with equilibrium abundances. In the case of \object{LB\,1919}, the
use of these new models resulted, however, in a worse fit compared to the
analysis of \citet{1998AuAfortable...329.1045W} with homogeneous models because
of too much levitation of metals in the diffusion models. The
reason for this overprediction remained unexplained.

\subsection{\textsl{FUSE} \label{subsect:LB1919FUSE}}

\begin{figure}
\centering \rotatebox{-90}{\includegraphics[width=.60\columnwidth]{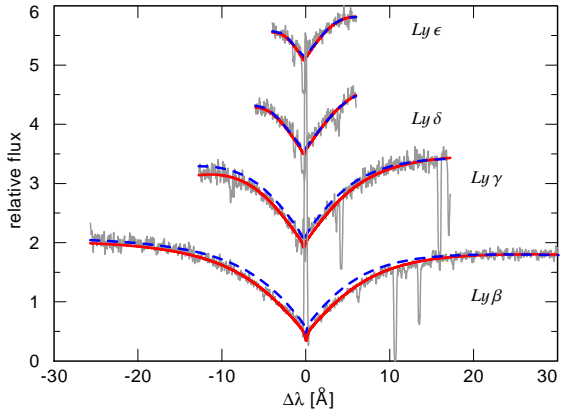}}
\caption{Fit to the Lyman lines of
  \object{LB\,1919} (thin line). Overplotted are pure H models with
  $\Teff=56\,000\,\text{K}$, $\logg=7.9$ (thick line), and with
  $\Teff=69\,000\,\text{K}$, $\logg=7.9$ (dashed line).
  \label{LB1919plotHmod_Lyman}}
\end{figure}

A grid of pure-hydrogen model atmospheres was used to fit the Lyman lines in the
\textsl{FUSE} spectral range. We found $\Teff = 56\,000 \pm 2000$\,K and
$\logg=7.9\pm 0.3$. Somewhat hotter models  with higher gravity (e.g., $\Teff =
62\,000\,K$, $\logg=8.2$) fit almost equally well, but are excluded by the
\textsl{Chandra} spectrum (Sect.\,\ref{sect:lbchandra}).  The temperature is
significantly lower than the Balmer-line results reported above. It is
remarkable that \citet{2007ApJ...667.1126L} arrived at a similarly low value of
57\,701$\pm$18\,400\,K from the overall UV (\textsl{IUE})/optical flux
distribution, and at an even lower value of 39\,045$\pm$18\,000\,K by fitting the
Ly$\alpha$ line in the \textsl{IUE} spectrum. Figure~\ref{LB1919plotHmod_Lyman}
shows the comparison between a hotter model and a better-fitting, cooler model
together with the \textsl{FUSE} Lyman lines. Adding metals to the 56\,000\,K
model with abundances at the upper limit derived below  has no significant
influence on the Lyman-line profiles.

Upper limits to metal abundances were determined with homogeneous model
atmospheres. Their accuracy is estimated to be about 0.3 dex following from the
error ranges in \Teff and \logg, and they are given in number ratios relative to
H. We also compare line profiles from diffusion models with observations.

\begin{figure}
\centering \rotatebox{-90}{\includegraphics[width=0.4\linewidth]{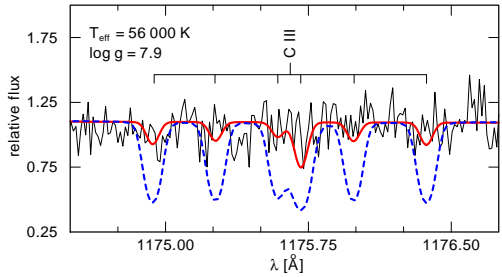}}
\caption{Homogeneous (thick line) and stratified (dashed line) model spectra
   near the C\,{\sc iii}
    multiplet at $\lambda\,1175$\,\AA\ in \object{LB\,1919}. The homogeneous
    model has C/H $=1.0\cdot 10^{-7}$. \label{LB1919_CIIItr_1174punkt93_5285}}
\end{figure}

\paragraph{Carbon and Oxygen}
From the absence of C\,{\sc iii} $\lambda\,1175$\,\AA, we derive an upper limit
of C/H = $1.0 \cdot 10^{-7}$. The stratified models overpredict the line
strengths (Fig.\,\ref{LB1919_CIIItr_1174punkt93_5285}). Note that all computed
line profiles have zero rotation velocity.
 
Except for the O\,{\sc vi} $\lambda\lambda$\,1031.91, 1037.61\,\AA\ resonance
doublet, no O line is  identified. Because of a possible ISM contribution, only
an upper limit of O/H $=1\cdot 10^{-6}$ can be derived. The stratified models
fit the observation best for $\logg=8.2$. A smaller gravity results in too
strong features.

\begin{figure}
    \centering \rotatebox{-90}{\includegraphics[scale=0.67]{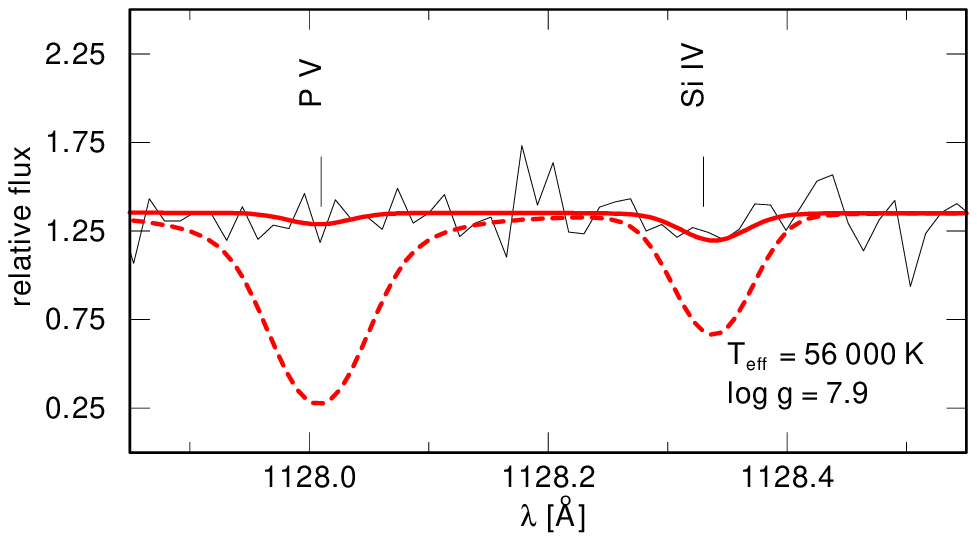}}
  \caption{Homogeneous (thick line) and stratified (dashed line) model profiles of P\,{\sc v}  and Si\,{\sc
      iv} compared to \object{LB\,1919}. The
    homogeneous model has Si/H $=5.0\cdot 10^{-9}$ and P/H $=1.0\cdot 10^{-10}$.}
    \label{LB1919_PV_1128punkt01_SiIV_1128punkt33_5285_nolabely}
\end{figure}

\begin{figure}
\centering \rotatebox{-90}{\includegraphics[width=0.4\linewidth]{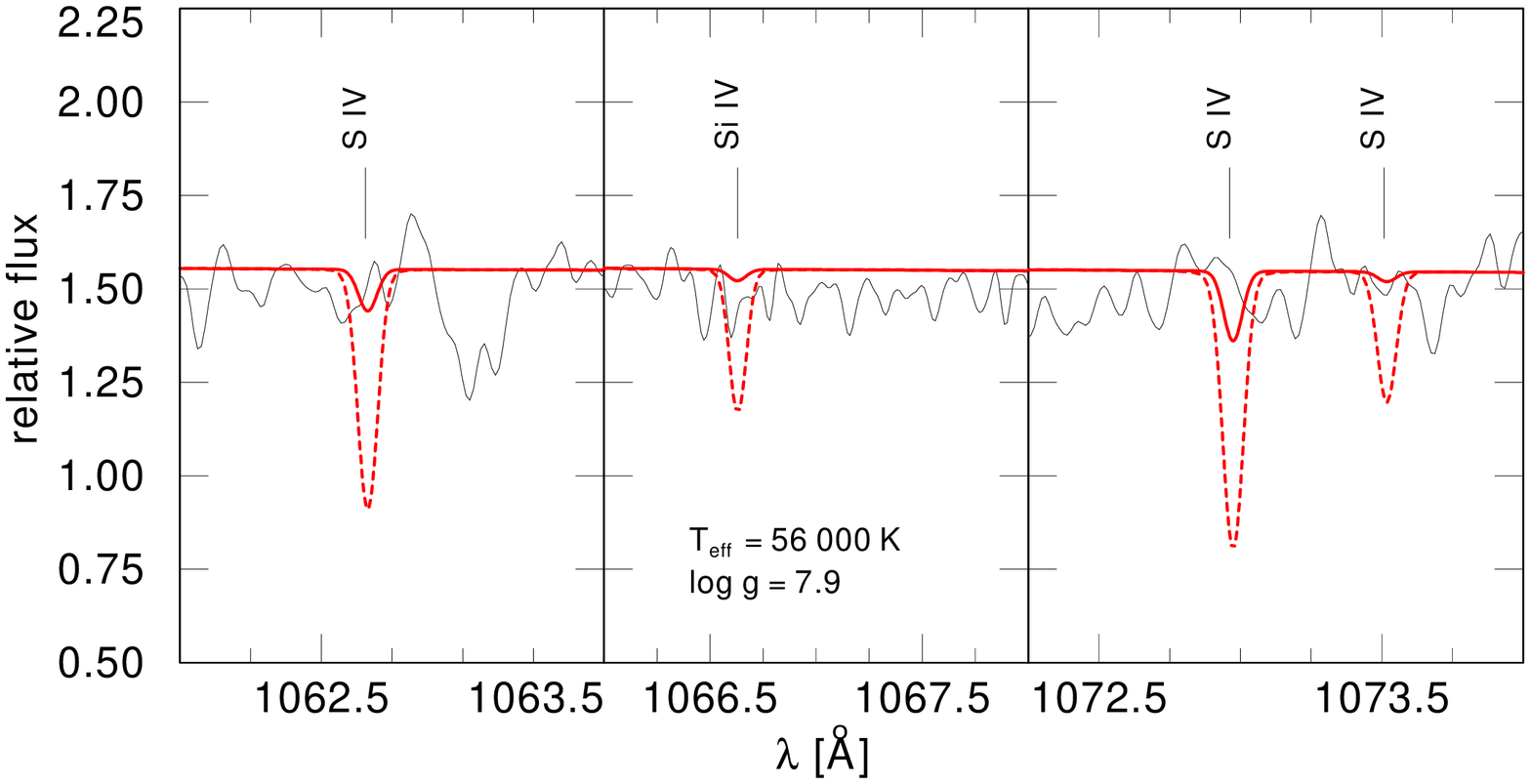}}
\caption{
Homogeneous (thick line) and stratified (dashed line) model profiles of
S\,{\sc iv} and
  Si\,{\sc iv} compared to \object{LB\,1919}. 
The homogeneous model has S/H $=1.75\cdot 10^{-8}$ and
  Si/H $=5.0\cdot 10^{-9}$. \label{LB1919_SIV_SiIV_5685}}
\end{figure}

\paragraph{Silicon, Phosphorus and Sulfur}
From the absence of Si\,{\sc iv}  $\lambda\lambda$\,1066.63, 1122.48,
1128.33\,\AA, P\,{\sc v} $\lambda\lambda$\,1117.98, 1128.01\,\AA, and S\,{\sc
iv} $\lambda\lambda$\,1062.66, 1072.96, 1073.508\,\AA, upper limits of
Si/H $=5.0 \cdot 10^{-9}$, P/H $=1.0 \cdot 10^{-10}$ and S/H $=1.75 \cdot
10^{-8}$ are derived
(Figs.\,\ref{LB1919_PV_1128punkt01_SiIV_1128punkt33_5285_nolabely},
\ref{LB1919_SIV_SiIV_5685}). In all cases, the diffusion models predict too
strong line profiles.

\paragraph{Iron and Nickel}
No lines are identified corresponding to upper limits of
$\text{Fe}/\text{H}<1.0\cdot 10^{-6}$ and $\text{Ni}/\text{H}<1.0\cdot
10^{-6}$. At $\Teff=56\,000$\,K, the stratified models
predict no detectable lines only if the surface gravity is $\logg=8.2$ or higher.

\subsection{EUVE}

\begin{figure}
\centering \rotatebox{-90}{\includegraphics[width=0.65\linewidth]{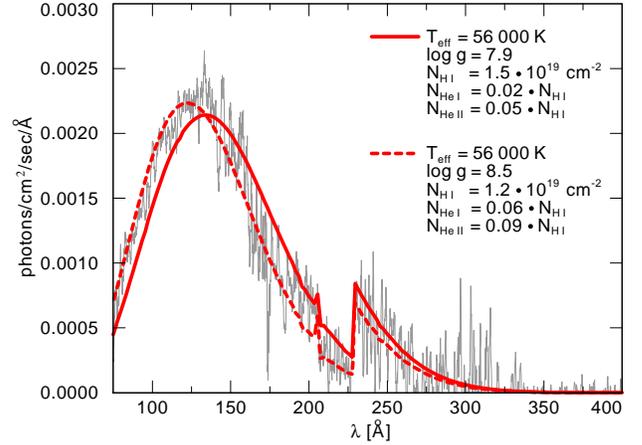}}
\caption{Best-fit homogeneous (solid) and stratified (dashed) models
  for the \textsl{EUVE} spectrum of
  \object{LB\,1919} (thin  line). The homogeneous model has metal abundances set
  to the upper limits given in Table~\ref{Abundance_table}; neither model has oxygen.\label{LB1919_euve_bestfit}}
\end{figure}

The stratified models calculated by \citet{2005PhDT.........1L} had a flux too
low to match the \textsl{EUVE} observation.  The failure to derive a good fit
resulted from the excessive opacity caused by the large number of elements
incorporated in the model atmospheres. We tried to reproduce the \textsl{EUVE}
observation with stratified models that include only those metals
listed in Table~\ref{Abundance_table}, i.e., metals that have been detected
in the \textsl{FUSE} spectra of comparable WDs, like GD\,246. Our results show
that the strongest influence on the EUV SED comes from the
opacity of oxygen and that a satisfying fit can only be derived for a model
atmosphere without this species and at high gravity (\logg = 8.5).  We also computed
homogeneous models with the same elements and abundances set to the upper
limits derived from the \textsl{FUSE} analysis (again without oxygen) and also obtained a
good fit (Fig.\,\ref{LB1919_euve_bestfit}). The same result is obtained from a
pure H model. 

\subsection{\textsl{Chandra}}\label{sect:lbchandra}

No Fe or Ni lines are detected in the \textsl{Chandra} spectrum, giving
upper abundance limits of Fe/H and Ni/H $<1.25\cdot 10^{-7}$ derived from
homogeneous Fe+H and Ni+H models. These limits are almost one order of magnitude
more stringent than the \textsl{FUSE}-derived results. The observed overall SED can be
reproduced successfully with both homogeneous and stratified models. The
interstellar H and He column densities were kept fixed to the
\textsl{EUVE}-derived values. For the stratified models, however,
there is still enough Fe and Ni levitated
that lines should be detectable in the \textsl{Chandra} spectrum, even at a
higher gravity of $\logg=8.5$.

\begin{figure}
\centering
\rotatebox{-90}{\includegraphics[width=0.7\linewidth]{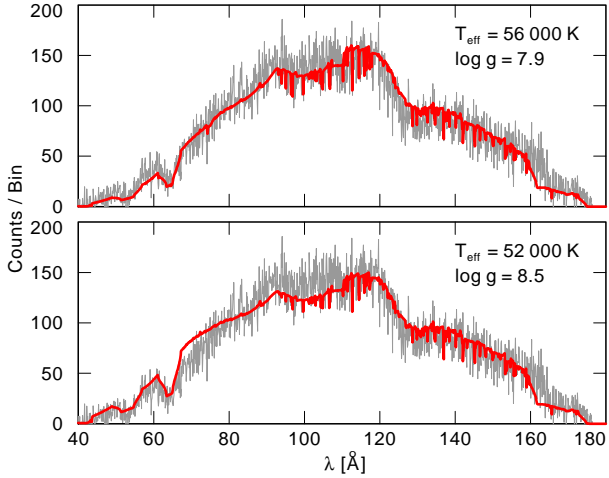}} 
\caption{Best-fit models (thick lines) for the \textsl{Chandra}
  spectrum of \object{LB\,1919} (thin lines). Top: homogeneous model, 
bottom: diffusion model. Both models are computed
  with H, C, O, Si, P, and
  S. The homogeneous model has abundances equal
  to the upper limits given in Table~\ref{Abundance_table}.\label{LB1919_Chandra_best_fit_5685_5285_unsoeld_Si}}       
\end{figure}

For further analysis, homogeneous and stratified models were calculated
that incorporated the same metals as the \textsl{EUVE} analyses in the preceding
section. Individual spectral features in the observation are not recognizable,
which serves as an additional constraint for the upper limits of the metal
abundances determined in the \textsl{FUSE} analysis. Both kinds of models match
the \textsl{Chandra} spectrum well. The models displayed in
Fig.\,\ref{LB1919_Chandra_best_fit_5685_5285_unsoeld_Si} include oxygen, the
homogeneous model with an abundance according to the upper limit given in
Table~\ref{Abundance_table}. The vast majority of the weak lines in these models
stems from O. In essence, a pure H model gives an equally good fit.

\subsection{Mass and distance}

To determine the mass from \Teff and \logg, we use DA evolutionary tracks by
\citet{1998MNRAS.296..206A}. We chose the tracks with metallicity $z=10^{-3}$
and hydrogen-envelope fractional mass of $\text{M}_{\text{H}}/ \text{M}_{\star}=
10^{-4}$ and derive $M=0.66\,^{+0.14}_{-0.12}\,\Msun$. Taking the flux
calibration of \citet{1984A&A...130..119H}, we compute the distance, using the
extinction-corrected visual magnitude that is determined from the observed
magnitude V=16.41 to $m_{V_0}={\rm V}-2.174 \cdot$ E(B$-$V) with E(B$-$V) =
0.030.  With $\Teff=56\,000\,\pm 2000$\,K, $\logg=7.9\pm0.3$, and
$H_{\mathrm{\nu}}=7.33\cdot 10^{-4}\,\text{erg cm}^{-2}\, \text{sec}^{-1}\,
\text{Hz}^{-1}$  (model Eddington flux at 5454\,\AA) we find (see,
e.g., \citealt{1994A&A...286..543R})
\begin{align}\label{eqDistance}
d=7.11\cdot 10^4 \sqrt{H_{\mathrm{\nu}}\, M\, 10^{0.4
m_{V_0}-\logg}}=321\,^{+101}_{-137}\,\text{pc}.
\end{align}

\subsection{Summary on \object{LB\,1919}}

Our fit to the Lyman lines and the  \textsl{Chandra} spectrum gave $\Teff=56\,000 \pm 2\,000$\,K, $\logg=7.9 \pm
0.3$. The temperature is significantly lower ($\approx 10\,000\,K$) than that
derived from previous Balmer-line fits. This result is surprising because
\citet{2003MNRAS.344..562B} had established that Lyman-line analyses for DAs
exceeding \Teff$\approx 50\,000$\,K yield systematically
\emph{higher} temperatures than Balmer-line analyses, i.e., the opposite that we
found for LB\,1919. In their sample, however, there is also one exception to
this general trend. Very similar to LB\,1919, \object{PG\,1342+444} has \Teff =
66\,750\,K and 54\,308\,K derived from Balmer and Lyman lines, respectively (see
also \citealt{2002MNRAS.330..425B}).

Metal lines are not detected in the \textsl{FUSE} and \textsl{Chandra}
spectra. We derived upper abundance limits with homogeneous models. The
respective models fit the SEDs observed by \textsl{Chandra} and \textsl{EUVE}.
Stratified models, on the other hand, overpredict metal abundances. 
Reducing the abundances in these models to bring them into agreement with
observations requires an excessively high gravity (\logg$>8.5$), which is not
compatible with the Balmer and Lyman lines. In essence, this confirms the result
obtained from the \textsl{EUVE} analysis by \citet{2005PhDT.........1L}. All
observations are compatible with a pure H atmosphere.

\begin{table*}
\begin{center}
\small
\caption[\object{GD\,246} Parameter]{\Teff and \logg 
  of \object{GD\,246} as determined by previous analyses and the present work
  (see also Fig.\,\ref{GD246_results}).  \label{GD246_parameter_table}} 
\begin{tabular}{rccllll}
\toprule
No. & \Teff [K] & \logg [$cm/s^2$] & \multicolumn{3}{l}{Method and Model Atmospheres} & Authors\\
\midrule
1   & 60\,100   & 7.72   & LTE &Balmer            & H            & \citet{1997ApJ...480..714V}  \\
2   & 58\,700   & 7.81   & LTE &Balmer            & H            & \citet{1997ApJ...488..375F}  \\
3   & 59\,000   & 7.80   & LTE &\textsl{EUVE}     & H+metals          & \citet{1998AuAfortable...329.1045W} \\
4   & 53\,100   & 7.85   & NLTE &Balmer           & H            & \citet{1999ApJ...517..399N}  \\
5   & 54\,000   & 7.80   & LTE &Lyman             & H            & \citet{2001AuAfortable...373..674W} \\
6   & 56\,000   & 8.20   & NLTE &\textsl{EUVE}  & H+metals stratified& \citet{2002AuAfortable...382..164S} \\
7   & 52\,402   & 7.89   & NLTE &Lyman            & H+metals          & \citet{2003MNRAS.344..562B} \\
8   & 51\,300   & 7.91   & NLTE &Balmer           & H+metals          & \citet{2003MNRAS.344..562B} \\
9   & 54\,400   & 7.90   & NLTE &Balmer           & H            & \citet{2005ApJS..156...47L} \\
10  & 57\,007   & 7.82   & LTE &Balmer            & H            & \citet{2009AuAfortable...505..441K} \\
11  & 54\,150   & 8.00   & NLTE &Balmer           & H            & \citet{2010ApJ...714.1037L} \\
12  & 56\,160   & 7.98   & NLTE &Balmer           & H            & \citet{2011ApJ...743..138G} \\
13  & 57\,000   & 7.60   & NLTE &Lyman            & H            & this work                   \\
14  & 55\,000   & 7.90   & NLTE &\textsl{Chandra} & H+metals stratified & this work                 \\
\bottomrule
\end{tabular}
\normalsize
\end{center}
\end{table*}

\begin{figure}
  \centering
  \includegraphics[scale=0.7]{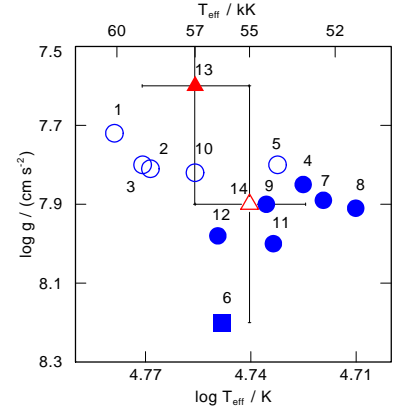}
  \caption{Results from previous analyses of GD\,246 
    (Table~\ref{GD246_parameter_table}). Results obtained by LTE
    analyses (open circles) are compared to NLTE analyses (filled
    circles) and the analysis with stratified NLTE model atmospheres
    (filled square). The triangles mark the two results of this paper, obtained by
    fitting the Lyman lines (full symbol) and the \textsl{Chandra} SED (open symbol).\label{GD246_results}}   
\end{figure}

\section{\object{GD\,246}}\label{sect:gd} 

\subsection{Previous investigations \label{previnvestGD246}}

Earlier determinations of \Teff and \logg of GD\,246  (\object{WD\,2309+105})
are summarized in Table~\ref{GD246_parameter_table} and
Fig.\,\ref{GD246_results}. The most recent study by \citet{2011ApJ...743..138G}
is probably the most reliable NLTE Balmer-line analysis because it utilizes new
Stark broadening data \citep{2009ApJ...696.1755T}. (We use the same data in our analysis.) All
analyses are based on Balmer- and/or Lyman-line profiles with the exception of
two, which are based on the \textsl{EUVE} SED. One of them
\citep{2002AuAfortable...382..164S} uses stratified NLTE models, and it results
in an exceptionally large gravity.

A small number of lines of trace metals could be identified in the \textsl{FUSE}
and \textsl{HST} spectra  of \object{GD\,246}. Beside elements known to appear
in WDs like Si, C, and P,  \citet{2005ApJ...622L.121V} identified for the first
time lines of germanium (Ge\,{\sc iv}) in a \textsl{HST} observation and
determined $\log \text{(Ge/H)}=-8.6 \pm 0.2$.

\subsection{\textsl{FUSE} and \textsl{HST}}

\begin{figure}
\centering \rotatebox{-90}{\includegraphics[width=.60\columnwidth]{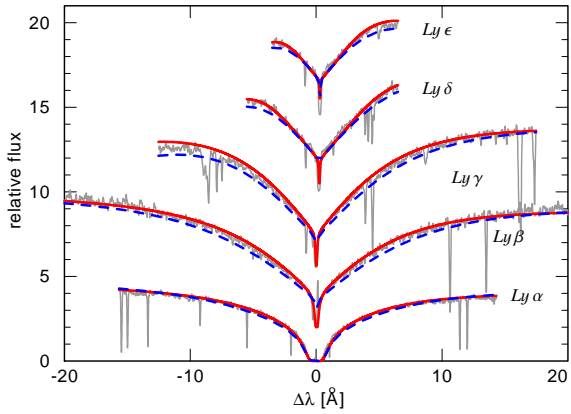}}
\caption{Fit to the Lyman lines of
  \object{GD\,246} (thin line). Overplotted are pure H models with
  $\Teff=57\,000\,\text{K}$, $\logg=7.6$ (thick line), and with
  $\Teff=52\,000\,\text{K}$, $\logg=7.9$ (dashed line).
  \label{GD246plotHmod_Lyman}}
\end{figure}

Pure hydrogen models were used to fit the Lyman lines. We found \Teff =
$57\,000\, \pm 2000$\,K and $\logg = 7.6 \pm 0.3$
(Fig.\,\ref{GD246plotHmod_Lyman}). Also plotted are the line profiles of one of
our models that has parameters that are essentially equal to those found by
\citet{2003MNRAS.344..562B} from their Lyman-line analysis
($\Teff=52\,000\,\text{K}$, $\logg=7.9$). It can be seen that the differences
between the two sets of profiles are rather subtle and the fit quality is
similar. We verified for some model parameters that the influence of metal-line
blanketing on the line profiles is negligible.

We then determined metal abundances. Some of them can be compared to the results
by other authors (Table~\ref{Abundance_table}).

\paragraph{Carbon}
No  C line is detectable. From the absence of the C\,{\sc iii}
$\lambda\,1175$\,\AA\ multiplet, we find an upper limit of C/H = $3.2\cdot
10^{-8}$. All stratified models predict strong C lines, indicating that they
clearly overpredict the C abundance.

\paragraph{Oxygen}
The result for O is similar. We derive an upper limit of O/H = $6\cdot 10^{-8}$
from the absence of a photospheric O\,{\sc vi} resonance doublet. The stratified
models strongly overpredict O.

\begin{figure}
\centering
\rotatebox{-90}{\includegraphics[scale=0.7]{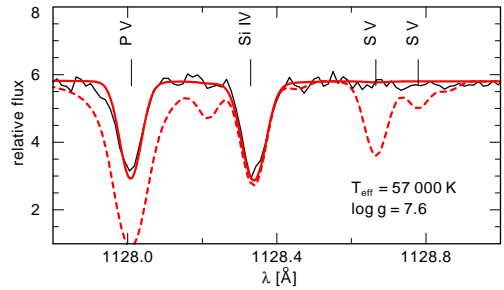}}  
\caption{Fit to P\,{\sc v} and Si\,{\sc iv} lines of \object{GD\,246} with
  a homogeneous (thick line) and a
  diffusion model (dashed line). The abundances in the homogeneous model are 
  Si/H = $1.2 \cdot 10^{-7}$, P/H = $3 \cdot
  10^{-9}$, and S/H = $5 \cdot 10^{-9}$.}
  \label{GD246_PV_1128punkt01_SiIV_1128punkt33_5579}
\end{figure}

\paragraph{Silicon} 
Si\,{\sc iv} lines are seen at $\lambda\lambda$\,1122.48, 1128.33\,\AA\
(Fig.\,\ref{GD246_PV_1128punkt01_SiIV_1128punkt33_5579}). We find Si/H =
$1.2(\pm 0.2)\cdot 10^{-7}$, that is identical to the result of
\citet{2003MNRAS.344..562B}. For the diffusion models, the lines fit at \Teff =
55\,000\,K and $\logg=7.9$.

\paragraph{Phosphorus}
Fitting the P\,{\sc v} $\lambda\lambda\, 1117.98, 1128.01$\,\AA\ resonance
doublet (Fig.\,\ref{GD246_PV_1128punkt01_SiIV_1128punkt33_5579}) gives P/H = $4
(\pm 2) \cdot 10^{-9}$. The stratified models predict too strong lines.

\paragraph{Sulfur}
From the absence of the S\,{\sc vi} $\lambda\lambda\,933.38, 944.52$\,\AA\
resonance doublet, we find S/H $ < 5 (\pm 2) \cdot 10^{-9}$. The stratified
models produce excessively strong lines
(Fig.\,\ref{GD246_PV_1128punkt01_SiIV_1128punkt33_5579}).

\paragraph{Iron and Nickel}
Fe and Ni lines are not detected in the UV range. We derive upper limits of
$1\cdot 10^{-6}$ for both species. Fe and Ni are overpredicted by the stratified
models.

\begin{figure}
\centering
\rotatebox{-90}{\includegraphics[width=0.4\linewidth]{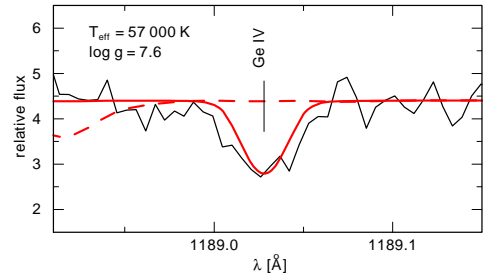}}
\caption{Fit to Ge\,{\sc iv} $\lambda\, 1189.03$\,\AA\ in
  \object{GD\,246} (thin line). Overplotted are a homogeneous model
  (thick line, Ge/H $=5\cdot 10^{-9}$) and a stratified 
  model (dashed line). \label{GD246_GeIV_1189.028_5573}}     
\end{figure}

\paragraph{Germanium}
Our fit to the Ge\,{\sc iv} $\lambda\lambda 1188.99, 1229.81$\,\AA\ resonance
doublet gives Ge/H $=5 (\pm 1)\cdot 10^{-9}$
(Fig.\,\ref{GD246_GeIV_1189.028_5573}). This is about twice the value derived by
\citet{2005ApJ...622L.121V} for a 56\,000\,K model. The difference can be traced
back to NLTE effects (Sect.\,\ref{subsectionGemodelatom}). The diffusion models
drastically underpredict the line strengths. At $\Teff=57\,000$\,K and
$\logg=7.6$ no Ge\,{\sc iv} lines are visible.

\begin{figure}
\centering
\rotatebox{-90}{\includegraphics[width=1.0\linewidth]{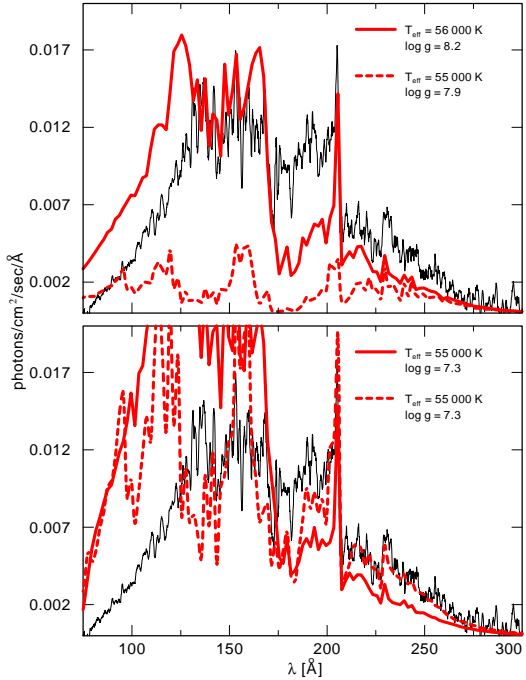}}  
\caption[GD246 \textsl{EUVE} best fit models]{Best-fit models to
  the \textsl{EUVE} spectrum of GD\,246 (thin line). 
Top: stratified models; bottom: homogeneous models. Solid lines: models
including Fe; dashed lines: models include Fe plus Ni.
\label{GD246_euve_bestfit}}     
\end{figure}

\begin{table*}
\begin{center}
\begin{threeparttable}
\small
\caption[]{Element
  abundances in  \object{LB\,1919} and
  \object{GD\,246} as determined from UV spectra
  with homogeneous models with $56\,000 \pm 2000$\,K and $\logg=7.9 \pm
  0.3$ for \object{LB\,1919} and $57\,000 \pm 2000$\,K and $\logg=7.6 \pm 0.3$
  for \object{GD\,246}. Identified ions are
  noted in the fifth column. Column six indicates the tendency
  of the abundance of a diffusion model to be stronger ($>$) or weaker
  ($<$) than the observed line strength. Literature values for the abundances
  are given in the last column. \label{Abundance_table}}   
\begin{tabular}{llrrlcr}
\toprule

WD       & Element    &  Abundance (X/H) & Uncertainty   & Ions
& Diff. & Literature\\
\midrule
\object{LB\,1919} \hspace{1mm}  & C & \hspace{15mm} $<1.0\cdot10^{-7}$ &
\hspace{7mm} $5.0\cdot10^{-8}$  & C\,{\sc iii}& $>$ &\\
          & O  & $<1.0\cdot10^{-6}$   &$2.0\cdot10^{-6}$& O\,{\sc vi}        & $>$ &\\
          & Si & $<5.0\cdot10^{-9}$   &$5.0\cdot10^{-9}$& Si\,{\sc iii+iv}  \hspace{2mm}& $>$  &\\
          & P  & $<1.0\cdot10^{-10}$  &$1.5\cdot10^{-10}$&P\,{\sc v}& $>$  &\\
          & S  & $<1.75\cdot10^{-8}$  &$7.5\cdot10^{-9}$ &S\,{\sc iv+vi}& $>$ &\\
          & Fe & $<1.0\cdot10^{-6}$   &$1.0\cdot10^{-6}$ &Fe\,{\sc v}& $>$&\\
          & Ni & $<1.0\cdot10^{-6}$   &$1.0\cdot10^{-6}$ &Ni\,{\sc v}& $>$&\\
\noalign{\smallskip}
\object{GD\,246}    & C  & $<3.2\cdot10^{-8}$ &$1.5\cdot10^{-8}$& C\,{\sc iii}       & $>$ &\\
          & O  & $<6.0\cdot10^{-8}$    & $5.0\cdot10^{-8}$ & O\,{\sc vi}
& $>$  & $1.6\cdot10^{-7}$ \tnote{a}\\
          & Si & $1.2\cdot10^{-7}$   & $2.0\cdot10^{-8}$   & Si\,{\sc iv}
& $=$  &$5.0\cdot10^{-8}$ \tnote{b} \\
          &    &                    &                   &           &
& $3.2\cdot10^{-8}$ \tnote{c}\\
          &    &                    &                   &           &
& $1.2\cdot10^{-7}$ \tnote{a}\\
          & P  & $3.0\cdot10^{-9}$     & $2.0\cdot10^{-9}$   & P\,{\sc v}
& $>$  & $6.3\cdot10^{-9}$ \tnote{c}\\
          &   &                     &                   &
&      & $7.5\cdot10^{-9}$ \tnote{b}\\
          & S  & $5.0\cdot10^{-9}$     & $2.0\cdot10^{-9}$   & S\,{\sc vi}
& $>$  & $<3.0\cdot10^{-7}$ \tnote{b}\\
          & Ge & $5.0\cdot10^{-9}$     & $1.0\cdot10^{-9}$   & Ge\,{\sc iv}
& $<$  & $2.5\cdot10^{-9}$ \tnote{d}\\
          & Fe & $<1.0\cdot10^{-6}$    & $1.0\cdot10^{-6}$ &Fe\,{\sc v} & $>$  & $<2.0\cdot10^{-5}$ \tnote{b}\\
          &    &                      &                     & & $>$  & $<1.2\cdot10^{-7}$ \tnote{a}\\
          & Ni & $<1.0\cdot10^{-6}$    & $1.0\cdot10^{-6}$  &Ni\,{\sc v}& $>$  & $<1.3\cdot10^{-7}$ \tnote{a}\\
\bottomrule
\end{tabular}
\begin{tablenotes}[para]\footnotesize
\item[a] \citet{2003MNRAS.341..870B}
\item[b] \citet{2001AuAfortable...373..674W}
\item[c] \citet{2001ASPC..226...90C}
\item[d] \citet{2005ApJ...622L.121V}
\end{tablenotes}
\normalsize
\end{threeparttable}
\end{center}
\end{table*} 

\subsection{\textsl{EUVE}}

Two types of models were calculated. Both include the metals identified in the
\textsl{FUSE} observation (the homogeneous models with the determined
abundances, and C and O set to their upper limits). For one type of models, Fe
is added, and the other type includes Fe plus Ni (setting Fe/H = $1.25\cdot
10^{-6}$ and Ni/H = $7.9\cdot 10^{-8}$ in the homogeneous models).
Fig.\,\ref{GD246_euve_bestfit} shows that the stratified models with Fe as
heaviest element provide a better fit to the observation than the models
also including Ni. A gravity of $\logg=8.2$ is needed, which is significantly
higher than the result of the Lyman-line fits. The homogeneous models fail
completely in reproducing the \textsl{EUVE} SED, whis is in agreement with the results
of \citet{2005PhDT.........1L}.

\subsection{\textsl{Chandra}\label{GD246:Chandra}}

\begin{figure*}
\centering
\rotatebox{-90}{\includegraphics[width=0.4\linewidth]{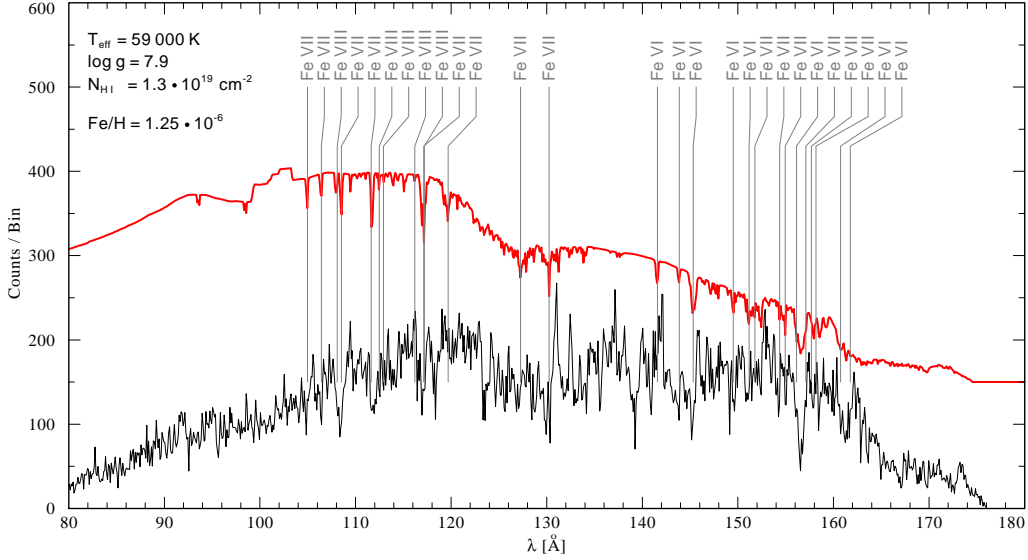}}
\caption{\textsl{Chandra} spectrum of \object{GD\,246} (thin line; bin
  size 0.1\,\AA) and model spectrum (shifted upward by 150 counts; Fe POS lines 
only). \label{Linien_Fe_chandra_pos_list_few}}
\end{figure*}

\citet{2002ASPC..262...57V} detected Fe lines in the \textsl{Chandra} spectrum
of \object{GD\,246}. They are identified in
Fig.\,\ref{Linien_Fe_chandra_pos_list_few}. Their NLTE analysis indicated
$\text{Fe/H}=3\cdot 10^{-7}$.

\begin{figure}
\centering
\rotatebox{-90}{\includegraphics[width=0.7\linewidth]{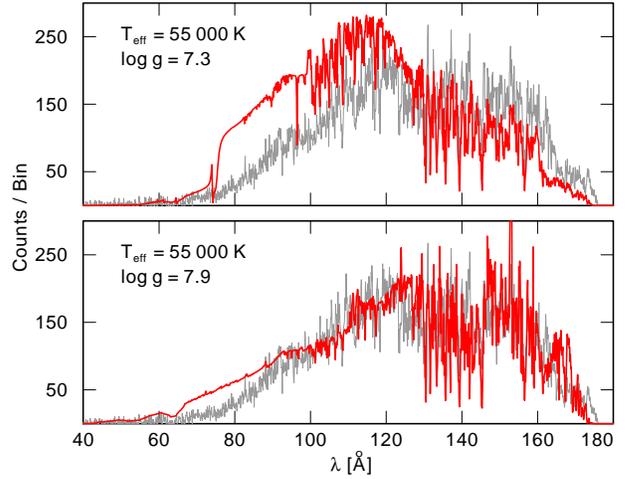}}
\caption[\object{GD\,246} \textsl{Chandra} best fit models]{Best-fit models
  (thick lines) to the \textsl{Chandra} spectrum of \object{GD\,246}. Top:
  Homogeneous model with abundances according to
  Table~\ref{Abundance_table}. Bottom: Stratified model. Both models have
  no Ni. \label{GD246_Chandra_best_fit_5573_5579_unsoeld}}    
\end{figure}

Fig.\,\ref{GD246_Chandra_best_fit_5573_5579_unsoeld} displays our best fits
using the homogeneous and stratified model types as employed for the
\textsl{EUVE} analysis. In general, as in the \textsl{EUVE} case, models
excluding Ni fit better. The final homogeneous model fits worse than the
Ni-truncated stratified model, and with an unrealistically low gravity of \logg
= 7.3. The best-fit (Ni-truncated) stratified model gives \Teff and \logg
that is in accordance with the most recent NLTE Balmer-line analysis of
\citet{2011ApJ...743..138G}; see Fig.\,\ref{GD246_results}. From the homogeneous
models, we confirm the Fe abundance derived by  \citet{2002ASPC..262...57V}.

\subsection{Mass and distance}

With \Teff $=57\,000 \pm 2000$\,K, $\logg=7.6\pm0.3$, $H_\nu=7.52\cdot
10^{-4}\,\text{erg cm}^{-2}\, \text{sec}^{-1}\, \text{Hz}^{-1}$,
$m_{\mathrm{v}}=13.09$, E(B$-$V) = 0.0011, we derive $M=0.54\,^{+0.12}_{-0.09}\,\Msun$
and $d=94^{+29}_{-40}$\,pc.

\subsection{Summary on \object{GD\,246}}

Our Lyman-line fit gave \Teff = 57\,000\,K and \logg = 7.6. The Balmer-line fit
by \citet{2011ApJ...743..138G} resulted in a similar temperature (56\,160\,K)
but at a significantly higher gravity (7.98). With our parameters, the
stratified model compared to the \textsl{FUSE} data overpredicts C, O, Fe,
  Ni, P,
and S, but underpredicts Ge. That result remains essentially valid for
stratified models with increased gravity (\logg = 7.9).

No acceptable fit to the \textsl{EUVE} spectrum was achieved. The gravity of the best fit
homogeneous model is too low (\logg = 7.3), while that of the
best-fit stratified model is too high (\logg = 8.2). It must be stressed that in
both cases the formally best fits are very poor.

For fitting the \textsl{Chandra} spectrum, the stratified models are only
better than the homogeneous models when  Ni is arbitrarily removed. The best-fit
stratified model has then \logg = 7.9, which is closer to the Balmer-line than
to the Lyman-line result. However, the fits deteriorate when Ni is
included. But even when Ni is excluded, the best-fit stratified model is not
satisfactory. We conclude that stratified models do not predict correctly
the individual element abundances. This conclusion is corroborated by the
\textsl{FUSE} analysis.

\begin{figure}
\centering
\rotatebox{-90}{\includegraphics[width=.7\linewidth]{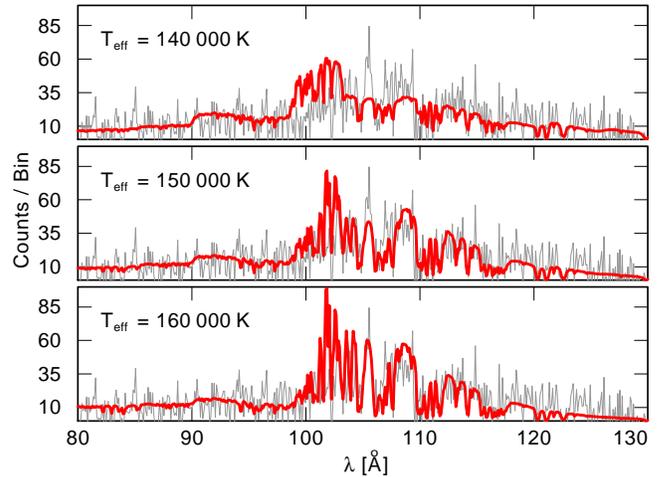}}
\caption{
Three models with different \Teff fitted to the \textsl{Chandra} spectrum of
PG\,1520+525. Model abundances are given in Table~\ref{ParameterPG1520}, \logg =
7.5, $\text{log N(H\,{\sc i})}=20.0$. \label{PG1520_140_150_160_75_vlg_neu}}         
\end{figure}

\begin{figure}
\centering
\includegraphics[width=1.\linewidth]{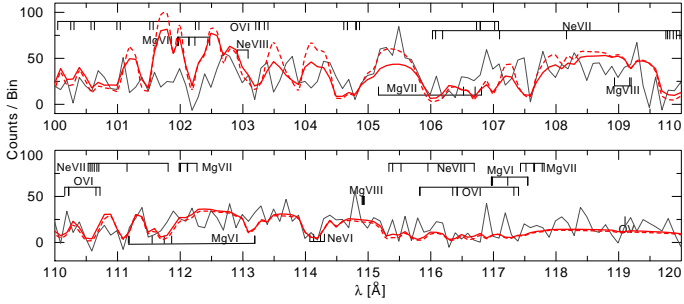}
\caption{Line identifications in 
  \object{PG\,1520+525}. Overplotted are two models with different \Teff:
  150\,000\,\text{K} (thick line), 160\,000\,\text{K} (dashed line). Model
  abundances 
are given in Table~\ref{ParameterPG1520}, \logg =
7.5, $\text{log N(H\,{\sc i})}=20.0$. \label{PG1520_chandra_1575_1675_nh20}}      
\end{figure}

\section{\object{PG\,1520+525}}\label{sect:pg} 

\subsection{Previous investigations} 

The first analysis of \object{PG\,1520+525} (\object{WD\,1520+525}) based on
optical spectra was performed by \citet{1991A&A...244..437W} and gave \Teff =
$140\,000 \pm 20\,000$\,K and \logg = $7.0 \pm 1.0$. A first indication that the
temperature exceeds 140\,000\,K was derived from the shape of an \textsl{EUVE}
spectrum \citep{1996aeu..conf..229W}.  Utilizing UV spectra,
\citet{1998A&A...334..618D} found \Teff = $150\,000 \pm 10\,000$\,K and \logg =
$7.5 \pm 0.5$. These parameters, as well as the He, C, O, and Ne abundances
\citep{2004A&A...427..685W}, are given in Table~\ref{ParameterPG1520}.

\subsection{\textsl{Chandra} analysis}

Most counts in the \textsl{Chandra} spectrum are located in the
100--120\,\AA\ region. At longer wavelengths, the ISM absorbs the photospheric
flux, while at shorter wavelengths, the flux decreases due to photospheric O\,{\sc
vi} bf absorptions. Although the S/N ratio is rather poor, it is possible to
identify individual lines from O\,{\sc vi} and Ne\,{\sc vii}
(Fig.\,\ref{PG1520_chandra_1575_1675_nh20}). The identification of Mg lines is
ambiguous. For a detailed list of possible lines in this wavelength range we
refer to the analysis of H\,1504+65 \citep{2004A&A...421.1169W}.

We have calculated a small grid of models for \Teff = 140\,000\,K, 150\,000\,K,
160\,000\,K, and \logg = 6.4, 7.0, 7.5. We kept fixed the He/C/O/Ne abundance
ratio but varied the Mg abundance from solar to ten times solar in order to
derive an upper limit. 

The fit is hardly affected by the particular choice of log\,$g$. Instead, \Teff
is the most sensitive parameter. Mainly judging from the spectral slope, we find
the best fit at \Teff = $150\,000 \pm 10\,000$\,K adopting \logg = $7.5
\pm 0.5$ (Fig.\,\ref{PG1520_chandra_1575_1675_nh20}). This confirms the
previous temperature determination from UV/optical results, but the
\textsl{Chandra} spectrum allows no further reduction of the error bar.

\begin{figure}
\centering
\rotatebox{-90}{\includegraphics[width=.30\linewidth]{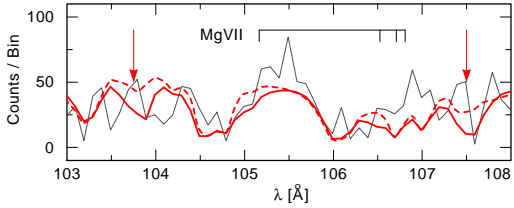}}
\caption{Detail of the \textsl{Chandra} observation of \object{PG\,1520+525} (thin
  line), overplotted with two models with solar (dashed line) and ten times solar (thick
  line) Mg abundance (\Teff=150\,000\,K, \logg = 7.5). \label{PG1520_chandra_Mg_1}}       
\end{figure}

A change in the Mg abundance from solar to ten times solar affects the strength
of the lines only marginally, and it is not clear which abundance is more likely
to fit the observation (Fig.\,\ref{PG1520_chandra_Mg_1}). The flux of the models
is also affected in regions where no Mg lines are expected. This can be
explained by the occurrence of autoionization features in the bf
cross sections. These resonances can be strong and narrow, looking like a common
absorption line \citep{2004A&A...421.1169W}. The wavelengths of these
resonances are uncertain and could thus be a possible explanation for
unidentified features in the observed spectrum.

\subsection{Summary on PG\,1520+525}

Our analysis of the \textsl{Chandra} spectrum gives an independent confirmation
of the result on \Teff derived from previous UV spectroscopy ($150\,000 \pm
10\,000$\,K, assuming \logg = $7.5 \pm 0.5$).

\begin{figure}
  \centering \includegraphics[scale=0.45]{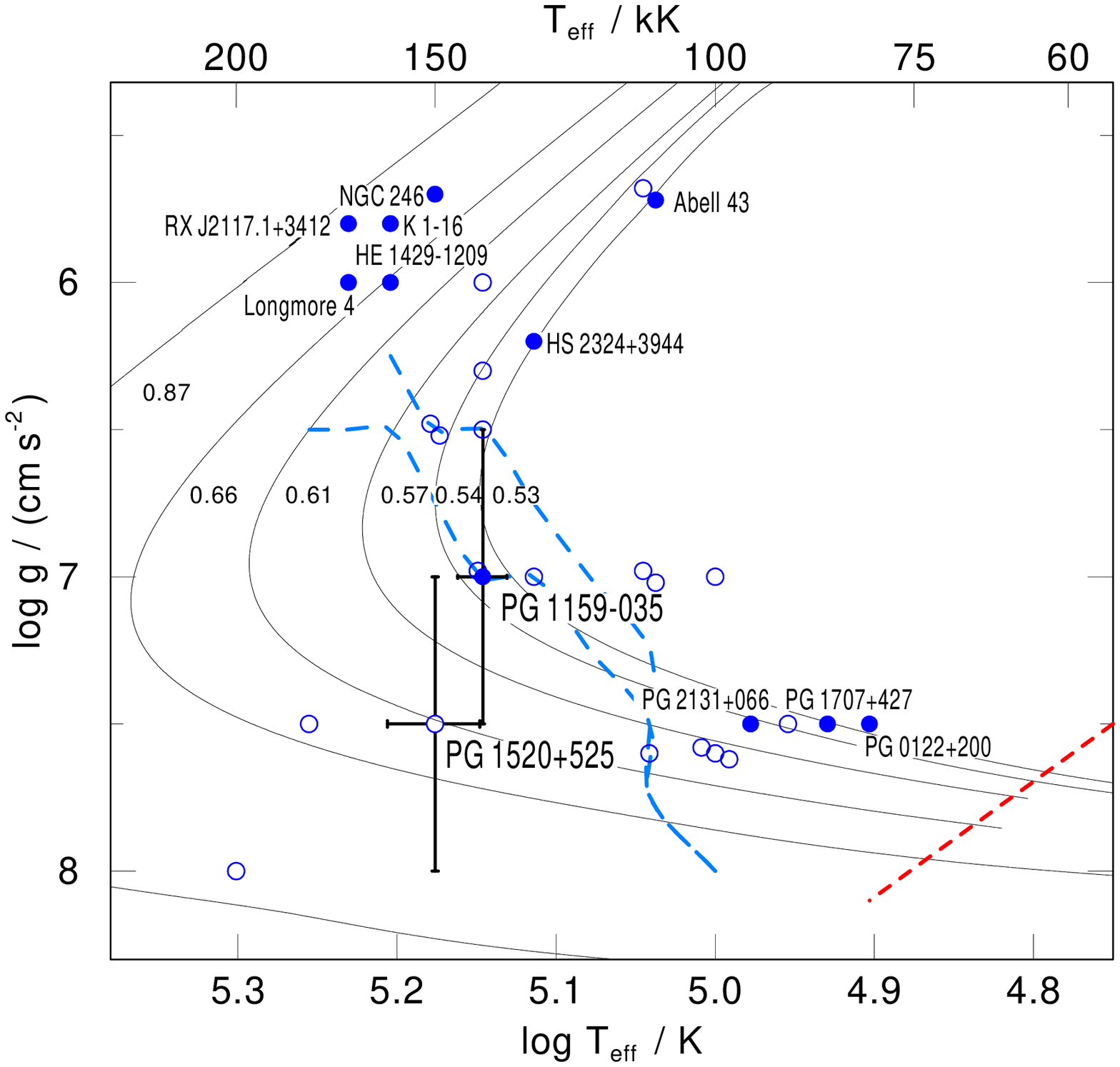}
  \caption{Pulsating (filled circles) and nonpulsating (empty
    circles) PG\,1159 stars in the $\log \Teff - \logg$ diagram. Evolutionary
    tracks are labeled with the respective stellar masses in M$_{\sun}$
    \citep{2006A&A...454..845M}. The red edge (short dashed line) of the
    instability region \citep{2006MmSAI..77...53Q} and two blue edges (long
    dashed lines,  \citealt{0067-0049-171-1-219}) are shown. The upper and lower
    blue edges are for $z=0$ and $0.007$, respectively.}
    \label{pulsators_inst_neu}
\end{figure}

As outlined in Sect.\,\ref{subsect:pgstars}, we may use this result to draw 
conclusions on the location of the blue edge of the GW~Vir instability region, which
is confined by the non-pulsator PG\,1520+525 and the pulsator PG\,1159$-$035.
The position of both stars in the \Teff --\logg diagram is displayed in
Fig.\,\ref{pulsators_inst_neu}. Also shown there is the theoretical location of
the blue edge predicted by \citet{0067-0049-171-1-219} assuming He/C/O mass
fractions of 0.4/0.4/0.2 in the stellar envelope, which is within error limits
identical to the abundances found in both stars
(Table~\ref{ParameterPG1520}). Besides these element abundances, the exact
location of the edge depends on the metallicity; therefore, it is shown here for
$z=0.0$ and $z=0.007$. The edge moves to higher effective temperatures with
increasing $z$. The $z=0.007$ line is appropriate for the two stars considered
here, because we had found solar Fe abundance in them
\citep{2011A&A...531A.146W}.  It can be seen that the theoretical position of
the edge runs between the location of both stars, very close to PG\,1159$-$035.

Also shown in Fig.\,\ref{pulsators_inst_neu} are other PG\,1159 stars in which
pulsations were looked for (taken from the compilation in
\citealt{2006PASP..118..183W}). 
The location of nonpulsators redward of the indicated blue edge is possibly due 
to their different chemical composition compared to
PG\,1159$-$035, because the composition affects the position of the instability
strip \citep{0067-0049-171-1-219}.

\begin{table}
\begin{center}
\small
\caption{Atmospheric parameters of \object{PG\,1520+525} and
  \object{PG\,1159--035}. \label{ParameterPG1520}}
\begin{tabular}{lll}
\toprule
          & \object{PG\,1520+525} & \object{PG\,1159--035} \\
\midrule
\Teff [K] & $150\,000\pm 10\,000$  & $140\,000\pm 5\,000$ \\
\logg(cgs)& $7.5\pm 0.5$          & $7.0\pm 0.5$      \\
He        & 0.43         & 0.33     \\
C         & 0.38         & 0.48     \\
O         & 0.17         & 0.17     \\
Ne        & 0.02         & 0.02     \\
Mg        & 0.006        & \\
Fe        & 0.0013       & 0.0013\\ 
\bottomrule
\end{tabular}
\begin{tablenotes}[para]\footnotesize
PG\,1520+525: \Teff, \logg\ from \citet{1998A&A...334..618D} and our work.\\ 
Abundances (in mass fractions) from \citet{2004A&A...427..685W},
except for \\ Mg (this work) and Fe \citep{2011A&A...531A.146W}. PG\,1159--035:
from \\ \citet{2011A&A...531A.146W}.
\end{tablenotes}
\normalsize
\end{center}
\end{table}

\section{Summary and Conclusions}\label{sect:sum} 

We analyzed the \textsl{Chandra} spectra of three hot WDs. Two of them
are H-rich WDs and the other one is a H-deficient PG\,1159
star.

\subsection{The DA white dwarfs \object{LB\,1919} and \object{GD\,246}}

The primary aim of our \textsl{Chandra} observation of LB\,1919 was to explain
its relatively low metallicity compared to similar objects as suggested by
previous analyses of \textsl{EUVE} spectra. It turned out that no metal features
were detected in the \textsl{Chandra} spectrum. The same result was found from
our analysis of a \textsl{FUSE} spectrum. In essence, all data are compatible
with the assumption that LB\,1919 has a pure hydrogen atmosphere. This is in
conflict with our stratified models. In them, the vertical run of individual
element abundances is computed from the assumption of equilibrium between
gravitational downward pull and radiative upward acceleration, i.e., the metal
abundances are not free parameters but computed as functions of \Teff\ and
\logg. They predict that significant amounts of light and heavy metals should
be accumulated and readily detectable in the atmosphere of LB\,1919.

A few other hot DAs with similar parameters that also do not have detected metals are
known \citep{2003MNRAS.341..870B}; one famous example is HZ\,43A (\Teff =
51\,000\,K, \logg = 7.9, \citealt{2006A&A...458..541B}). The reason for the
purity of their atmospheres is completely unknown. It may be speculated that
these peculiar DAs have no heavy-element reservoir that is assumed to be present
in the radiative levitation calculations. \citet{2003MNRAS.341..870B}
consider whether the possible depletion of those reservoirs by selective
mass-loss could be responsible for that phenomenon 
or whether the progenitors were metal-poor. They
argue that, for radiation-driven winds, mass loss should be lower in the pure H
stars than in those containing heavy elements, because the wind is driven by
metal lines. Furthermore, these WDs are local disc objects, so it seems
unlikely that any of the progenitors could have been metal poor.

We have also analyzed an archival \textsl{Chandra} observation of the hot DA
GD\,246. Our work was motivated by the fact that this is the only DA that
exhibits individual metal-line features in the soft X-ray range. As for
LB\,1919, we utilized chemically homogeneous as well as stratified
models. While we in principle expect that the stratified models are a better
representation, the \textsl{Chandra} observation cannot be fitted
satisfactorily by either type of model. A better fit by stratified models is
only achieved when nickel is removed artificially. Otherwise, the atmospheric
opacity becomes much too large because of a strong overprediction of nickel.
In comparison with the \textsl{FUSE} spectrum, most metals are overpredicted,
but at the same time, another one (Ge) is underpredicted. In the latter case, one
can speculate that the model atoms are still too small, not having enough line
transitions to absorb photon momentum. The overprediction of most metals,
however, is more of a problem. We can only speculate that an additional physical
mechanism, ignored by our models, is affecting the equilibrium abundances,
e.g., (selective) mass-loss or weak magnetic fields.

The complete failure of our model atmospheres, homogeneous or stratified, to fit
the \textsl{EUVE} observation reinforces the conclusion that some physics is
missing. The same conclusion was drawn by \citet{2003MNRAS.341..870B} from the
fact that UV spectra from DAs with similar \Teff\ and \logg\ display a wide
variety of metal abundances. These authors considered the possibility that
accretion of interstellar or circumstellar matter could cause this
variety. Recent investigations have revealed the frequent incidence of dust
debris disks around WDs (e.g., \citealt{2011AIPC.1331..193F}) that are composed of
disrupted planetary material.  In order to test the influence of accretion of
such material on the observed photospheric abundances, this effect would have to
be included in the diffusion/levitation models in a manner performed (e.g., by
\citealt{2009A&A...498..517K}) in cooler WDs in which radiative levitation is
negligible. However, it is difficult to understand how an external supply
of material would resolve the problem of present diffusion models
generally overpredicting metal abundances.

Our analysis of the two DAs comprised the derivation of effective temperature
and gravity from the Lyman lines, and the resulting parameters were compared to
published results from Balmer-line analyses. We confirm previous findings that
conflicting results are obtained from the UV and optical line profile
analyses. In particular, Balmer-line temperatures are often significantly lower than
Lyman-line temperatures \citep{2003MNRAS.344..562B}. We found
that LB\,1919 is another rare case where the opposite effect was
found. While this obviously points to a shortcoming in the WD atmosphere models,
there is no indication as to which physical ingredient is treated
inadequately or is missing. These uncertainties in the derived atmospheric
parameters add to the problems in a detailed quantitative comparison,
element by element, of observed metal abundances with predictions from radiative
levitation models.

\subsection{The PG\,1159 star \object{PG\,1520+525}}

The \textsl{Chandra} spectrum of the nonpulsator PG\,1520+525 was used to
constrain its effective temperature and to compare it with that of the pulsator
PG\,1159$-$035. They both confine empirically the blue edge of the GW~Vir
instability region for a particular envelope composition. The position of the
edge predicted by the nonadiabatic pulsation models of
\citet{0067-0049-171-1-219} is consistent with the spectroscopic results.
This is strong proof for the
predictive power of these  models. The interior structure of PG\,1159 stars that
is inferred by their usage appears to be very reliable, so that corresponding
asteroseismologic analyses are based on solid ground.

\begin{acknowledgements} 
We thank the referee for a thorough and constructive report.
JA and TR were supported by the German Research Foundation (DFG) under grant
WE\,1312/39-1 and the German Aerospace Center (DLR) under grant 05\,OR\,0806,
respectively. JJD was supported by NASA contract NAS8-39073 to the {\it
Chandra X-ray Center} during the course of this work. This research has made use of the SIMBAD database, operated at
CDS, Strasbourg, France.  Some of the data presented in this paper were obtained
from the Mikulski Archive for Space Telescopes (MAST). STScI is operated by the
Association of Universities for Research in Astronomy, Inc., under NASA contract
NAS5-26555. Support for MAST for non-HST data is provided by the NASA Office of
Space Science via grant NNX09AF08G and by other grants and contracts.
\end{acknowledgements}

\bibliographystyle{aa} \bibliography{LB1919_AA_bib}

\end{document}